\documentclass[11pt]{article}

\usepackage[text={6.8in,8.5in},centering]{geometry}
\usepackage{times}
\usepackage{amsthm}
\usepackage{amsmath,amsfonts,amssymb} 
\usepackage{array,delarray,xspace} 
\usepackage[pdftex]{graphicx} 
\usepackage{dsfont} 
\usepackage{xspace} 
\usepackage{psfrag}
\usepackage{epsfig}
\usepackage{graphics}
\usepackage{epic}
\usepackage{eepic}
\usepackage{color}
\usepackage{wrapfig}
\usepackage{multirow}
\usepackage{url}
\usepackage{subfig}
\usepackage{hyperref}

\def\final{0} 
\ifnum\final=1  
\newcommand{\vnote}[1]{[{\small Vicky: \bf #1}]\marginpar{*}}
\newcommand{\sidecomment}[1]{\marginpar{\tiny #1}}
\else 
\newcommand{\vnote}[1]{}
\newcommand{\sidecomment}[1]{}
\fi  

\newtheorem{lemma}{Lemma}[section]
\newtheorem{theorem}[lemma]{Theorem}

\newcommand{\ms}[1]{\ensuremath{\mathsf{#1}}}
\newcommand{\bra}[1]{\ensuremath{\langle#1|}}
\newcommand{\ket}[1]{\ensuremath{|#1\rangle}}
\newcommand{\argmax}{\operatornamewithlimits{arg\ max}}

\renewcommand{\bar}{\overline}

\newcommand{\ver}{{\ms{V}}}
\newcommand{\edge}{{\ms{E}}}

\newcommand{\nbr}{{\ms{nbr}}}

\newcommand{\energy}{{\mathcal{E}}}
\newcommand{\eng}{{\mathcal{E}}}
\newcommand{\oy}{{\mathcal{Y}}}

\newcommand{\ham}{{\mathcal{H}}}
\newcommand{\mat}{{\mathcal{M}}}

\newcommand{\art}{{\ms{ART}}}
\newcommand{\EC}{Exact Cover }
\newcommand{\SEC}{{\sc{EC3 } }}

\newcommand{\SAT}{3SAT }

\newcommand{\desev}{{\sc DeSEV}}

\newcommand{\gmin}{g_{\ms{min}}}
\newcommand{\wmis}{{\ms{mis}}}

\newcommand{\GEC}{G_{\ms{EC}}}
\newcommand{\GSAT}{G_{\ms{SAT}}}
\newcommand{\GCK}{G}

\begin{document}

\title{Adiabatic Quantum Algorithms for the NP-Complete Maximum-Weight Independent Set, Exact Cover and 3SAT Problems}

\author{Vicky Choi 
\\{\em vchoi@cs.vt.edu}
\\ Department of Computer Science
\\ Virginia Tech
\\Falls Church, VA
 }

\maketitle

\begin{abstract}
The problem Hamiltonian of the adiabatic quantum algorithm for 
the maximum-weight independent set problem (MIS) that is based on the reduction to the Ising problem 
(as described in \cite{minor-embedding}) has flexible parameters.  
We show that by choosing the parameters appropriately in the problem Hamiltonian (without changing the problem to be solved)
for MIS on CK graphs~\cite{CK08}, we can prevent the first order quantum phase transition~\cite{AC09} and 
significantly change the minimum spectral gap.
We raise the basic question about what the appropriate formulation of adiabatic running time should be.
We also describe adiabatic quantum algorithms for \EC and \SAT  in which the problem Hamiltonians 
are based on the reduction to MIS.  We point out that the argument in Altshuler et al.~\cite{altshuler-2009}
that their adiabatic quantum algorithm
failed with high probability for randomly generated instances of \EC does not carry over to this new algorithm.
\end{abstract}

\section{Introduction}
Adiabatic quantum computation (AQC) was proposed by
Farhi~et~al.~\cite{FGGLLP01} in 2000 as an alternative quantum paradigm to solve NP-hard 
optimization problems, which are believed to be classically intractable.
Later, it was shown by Aharonov~et~al.~\cite{ADKLLR04} that AQC is not
just limited to optimization problems, and is polynomially equivalent to 
 conventional quantum computation (quantum circuit model). 
A quantum computer promises extraordinary power over a classical computer,  as demonstrated by 
Shor~\cite{shor} in 1994 with the polynomial quantum algorithm for solving the factoring problem, 
for which the best known classical algorithms are exponential.
Just how much more powerful are quantum computers? 
In particular, we are interested in whether 
an adiabatic quantum computer can
 solve NP-complete problems  
{\em more} efficiently than a classical computer. 

Unlike classical computation or  conventional quantum model 
in which an algorithm is specified by a 
finite sequence of {\em discrete} operations via classical/quantum gates, the adiabatic 
quantum algorithm is {\em continuous}.
It has been assumed (see Section~\ref{sec:AQA} for more discussion) that, 
according to the adiabatic theorem, 
the dominant factor of the adiabatic running time (\art) of the algorithm 
scales polynomially with the inverse
of the {\em minimum spectral gap} $\gmin$ of the system Hamiltonian (that describes the algorithm). 
Therefore, in order to 
 analyze the running time of
an adiabatic algorithm, it is necessary to be able to 
bound $g_{\ms{min}}$ analytically. 
However,  $\gmin$ is in general difficult to compute (it is as
hard as solving the original problem if computed directly).
Rigorous analytical analysis of adiabatic algorithms remains challenging.
Most of studies have to 
resort to numerical calculations.
These include
numerical integration of Schr\"odinger equation
\cite{FGGLLP01,childs-clique}, 
eigenvalue computation (or exact diagonization)\cite{znidaric-2005-71,symmetries}, and quantum Monte Carlo (QMC) technique
\cite{Young,young-2009}. 
However,  not only are these methods  limited to small sizes 
(as the simulations of
quantum systems grow exponentially with the system size), but also little insight can be gained from these  numbers
to design and analyze the time complexity of the algorithm.

Perhaps, from the algorithmic design point of view, it is more 
important to unveil the quantum evolution black-box
and thus enable us to obtain insight for designing
efficient  adiabatic quantum algorithms.
For this purpose, we devise a visualization tool, called Decomposed State Evolution Visualization (\desev).
Through the aid of this tool, we  constructed a family of instances of MIS,
called CK graphs~\cite{CK08}. The numerical results of an adiabatic algorithm for MIS on these graphs suggested
that $\gmin$ is exponentially small and thus the algorithm requires exponential time.  These results
were then explained by the first order quantum phase transition (FQPT) in \cite{AC09}. 
Since then, there have been some other papers (Altshuler
et al.,~\cite{altshuler-2009} ; Farhi et al., ~\cite{farhi-2009}; Young et al.,~\cite{young-2009}; Jorg et al.,~\cite{Jorg1,Jorg2})
 investigating the same phenomenon, i.e., first order
quantum phase transition.
In particular, Farhi~et~al. in~\cite{farhi-2009} suggested that the exponential small gap caused by the FQPT
 could be overcome (for the set of instances they consider) by randomizing the choice of initial Hamiltonian. 
In this paper, 
we show that by 
changing the parameters in the problem Hamiltonian (without changing the problem to be solved) of the adiabatic algorithm for 
MIS on CK graphs, we prevent the FQPT from occurring and significantly increase $\gmin$.
We do so by scaling the vertex-weight of the graph, namely, multiplying the weights of vertices by a scaling factor.
In order to determine the best scaling factor, we raise the basic question about what the appropriate formulation of adiabatic running time should be.

We also describe adiabatic quantum algorithms for \EC and \SAT  in which the problem Hamiltonians 
are based on the reduction to MIS. 
In~\cite{altshuler-2009},  Altshuler et al.~claimed  that a particular adiabatic quantum algorithm failed
with high probability for randomly generated instances of  Exact Cover.
They claimed that the correctness of their argument did not rely on the specific form of 
the problem Hamiltonian for Exact Cover.  
We demonstrate an adiabatic algorithm for \EC in which the problem Hamiltonian is based on the reduction to MIS that  
questions the generality of this claim.

This paper is organized as follows. 
In Section~\ref{sec:AQA}, we review the adiabatic quantum algorithm, and
adiabatic running time. 
In Section~\ref{sec:AA-mis}, we recall the adiabatic quantum algorithm for MIS based on the reduction to the
Ising problem.
In Section~\ref{sec:desev}, we describe the visualization tool \desev{} and the CK graphs. We show examples of 
\desev{} on the MIS adiabatic algorithm for CK graphs. 
In Section~\ref{sec:change-parameter}, we describe how changing the parameters affects $\gmin$, and raise the question about $\art$.
In Section~\ref{sec:ec-sat}, we describe adiabatic algorithms for \EC and \SAT that are based on MIS reduction.
We conclude with the discussion in Section~\ref{sec:discussion}.

\section{Adiabatic Quantum Algorithm}
\label{sec:AQA}
An adiabatic quantum algorithm is described by a time-dependent system Hamiltonian
\begin{equation*}
\ham(t) = (1-s(t))\ham_{\ms{init}} + s(t) \ham_{\ms{problem}}
\end{equation*}
for $t \in [0,T]$, $s(0)=0$, $s(T)=1$.
There are three components of $\ham(.)$: 
(1) initial Hamiltonian: $\ham(0)=\ham_{\ms{init}}$;
(2) problem Hamiltonian:  $\ham(T)=\ham_{\ms{problem}}$;
and (3) evolution path: $s : [0,T] \longrightarrow [0,1]$, e.g., $s(t)=\frac{t}{T}$.
$\ham(t)$ is an adiabatic algorithm for an optimization problem if we encode the problem into the problem 
Hamiltonian $\ham_{\ms{problem}}$ such that the ground state of $\ham_{\ms{problem}}$ corresponds to the answer to
the problem. The initial Hamiltonian $\ham_{\ms{init}}$ is chosen to be non-commutative with $\ham_{\ms{problem}}$
and its ground state must be known  and experimentally constructable, e.g.,  $\ham_{\ms{inital}} = -\sum_{i \in \ver(G)} \Delta_i \sigma_i^x$.
Here $T$ is the running time of the algorithm.
According to the adiabatic theorem, if $\ham(t)$ evolves ``slowly'' enough, or equivalently, if $T$ is ``large'' enough (see Adiabatic Running Time below)
 the system remains in the ground state of $\ham(t)$, and consequently, ground state of $\ham(T)=\ham_{\ms{problem}}$ gives the solution to the problem. 

Notice that given a problem, there are three components (initial Hamiltonian, problem Hamiltonian, and evolution path) that specify
an adiabatic algorithm for the problem. A change in an  component (e.g. initial Hamiltonian) will result in a different adiabatic algorithm for 
the same problem. 

In this paper,  we fix the evolution path by the linear interpolation function $s(t)=\frac{t}{T}$. 
Hereafter,  we describe an adiabatic algorithm by the re-parametrized  Hamiltonian
\begin{equation*}
\ham(s) = (1-s)\ham_{\ms{init}} + s \ham_{\ms{problem}}
\end{equation*}
where $s \in [0,1]$, with understanding that $s(t)=t/T$.
Furthermore, throughout this paper, we fix the initial Hamiltonian to be  $\ham_{\ms{init}} = - \sum_{i \in \ver(G)}  \sigma_i^x$.
When it is clear from context, we also refer to the problem Hamiltonian as the adiabatic algorithm for the problem.

\medskip
\noindent{\bf Adiabatic Running Time.}
In their original
work~\cite{FGGS00}, the running time of the adiabatic
algorithm is defined to be the same as the adiabatic evolution time
$T$, which is given by the adiabatic condition of 
the adiabatic theorem. 
However, this definition is under the assumption of some physical limit of the maximum energy of the system (see e.g.,~\cite{jordan1}),
and is not well-defined from the computational point of view, as observed by
Aharonov~et~al.~\cite{ADKLLR04}. 
They re-define $\art(\ham)$ as 
$T \cdot
\max_s||\ham(s)||$, taking into the account of the time-energy 
trade-off in the Schr\"odinger's equation\footnote{ 
Namely,
$
i\frac{d\ket{\psi(s)}}{ds} = T\cdot \ham(s) \ket{\psi(s)}
= \frac{T}{K}\cdot K \ham(s) \ket{\psi(s)} 
$
where $K>0$ is a constant.}.

On the other hand, given the extensive work on the rigorous proofs of the adiabatic theorem, 
it is interesting (if not confusing)  that many different versions of the
adiabatic conditions
have been recently proposed. 
These include 
 \cite{adt-1,adt-2,adt-3,adt-4,adt-5,adt-6,adt-7,adt-8,adt-9,adt-10,adt-11} in
    the quantum physics community, and
\cite{Reichardt-04,ADKLLR04,adt-AR} in
the computer science community.
Most of these studies suggest that
\art{} scales polynomially with the inverse of the spectral gap
of the system Hamiltonian, which is 
sufficient when one is interested in the coarse computational complexity
of algorithms, namely, the distinction between polynomial and
exponential running time.

However, from both the practical and algorithmic point of view, 
it is important to have a more precise formulation of \art. 
First, this is because the specification of the adiabatic evolution
time $T$ is required in an adiabatic algorithm, and
therefore a tight and simple upper bound is desired. 
Second, we are interested in the actual time complexity of the
algorithm, and not just the polynomial vs. exponential distinction.
It is necessary to have a more precise formulation of \art{} such that
basic algorithmic analysis can be carried out. 
Third, at this stage of research, 
it is particularly important to have such a formulation
because the spectral gap, which plays the dominating role in the
formulation of \art, is difficult to analyze. All current efforts
on the spectral gap analysis
resort to numerical studies, 
and that means the studies are
 restricted to small problem sizes only. Therefore, to gain 
insight into the time complexity of algorithms from these small
instances, it is important that the formulation of \art{} applies to small sizes.
So what is the appropriate formulation of \art? What should the adiabatic condition(s) be? 
In Section~\ref{sec:art}, we compare three closely related versions and 
raise the question about what the appropriate adiabatic running time should be.

\section{An Adiabatic Algorithm for MIS}
\label{sec:AA-mis}
In this section, we recall the adiabatic algorithm for MIS that is based on the reduction to the Ising problem, as described in \cite{minor-embedding}. 
First, we formally define the Maximum-Weight Independent Set (MIS)
problem (optimization version):

\smallskip
\hspace*{0.7cm}{\bf Input:} An undirected graph $G (=(\ver(G),\edge(G)))$, where each vertex $i \in \ver(G) = \{1, \ldots, n \}$ is weighted by a
positive rational number $c_i$

\hspace*{0.7cm}{\bf Output:} A subset $S \subseteq \ver(G)$ such that
$S$ is independent (i.e., for each $i,j \in \ver(G)$, $i\neq j$, $ij
\not \in \edge(G)$) and the total
{\em weight} of $S$ ($=\sum_{i \in S}
c_i$) is maximized. 
Denote the optimal set by $\wmis(G)$.
\smallskip


There is a one-one correspondence between the  MIS problem and the Ising
problem, which is the problem directly solved by the quantum processor
that implements 1/2-spin Ising Hamiltonian. We recall the
quadratic binary optimization formulation of the problem.
More details can be found in \cite{minor-embedding}.
\begin{theorem}[Theorem 5.1 in \cite{minor-embedding}]
If $J_{ij} \ge \min\{c_i,c_j\}$ for all $ij \in \edge(G)$, then the maximum
  value of
  \begin{equation}
\oy(x_1,\ldots, x_n) = \sum_{i \in \ver(G)}c_i x_i - \sum_{ij \in \edge(G)}
  J_{ij}x_ix_j
\label{eq:Y}
  \end{equation}
is the total weight of the MIS. 
In particular if $J_{ij} > \min\{c_i,c_j\}$ for all
      $ij \in \edge(G)$, then $\wmis(G) = \{i \in \ver(G) : x^*_i = 1\}$,
where $(x^*_1, \ldots, x^*_n) = \argmax_{(x_1, \ldots, x_n) \in \{0,1\}^n}
\oy(x_1, \ldots, x_n)$.
\label{thm:mis}
\end{theorem}

Here the function $\oy$ is called the pseudo-boolean function for MIS.
Notice that in this formulation, we only require $J_{ij} > \min\{c_i,c_j\}$, and thus there is freedom in
choosing this parameter. In this paper we will show how to take advantage of this.

By changing the variables ($x_i=\frac{1+s_i}{2}$), it is easy to show that MIS is equivalent
to minimizing the following function, known as the {\em Ising energy function}:
\begin{eqnarray}
  \energy(s_1, \ldots, s_n) &=& \sum_{i \in \ver(G)} h_i s_i + \sum_{ij \in \edge(G)} J_{ij}s_is_j,
\end{eqnarray}
which is the 
eigenfunction of the following
 {\em Ising Hamiltonian}:
\begin{equation}
\ham_{\ms{Ising}} = \sum_{i \in \ver(G)} h_i \sigma^z_i + \sum_{ij \in \edge(G)} J_{ij}
\sigma^z_i \sigma^z_j
\label{eq:Ising}
\end{equation}
where $h_i = \sum_{j \in \nbr(i)}
  J_{ij} - 2c_i$, $\nbr(i) =\{j: ij \in \edge(G)\}$,
for $i \in \ver(G)$.

That is, an adiabatic algorithm for MIS in which the problem Hamiltonian is $\ham_{\ms{Ising}}$ 
is described by the following 
system Hamiltonian:
\begin{equation*}
\ham(s) = (1-s)\ham_{\ms{init}} + s \ham_{\ms{Ising}}
\end{equation*}
 where $s \in [0,1]$ with the assumption that $s(t)=t/T$.
If $T$ is sufficiently large according to the adiabatic theorem, then 
the ground state of $\ham(1)$, say $\ket{x_1^*x_2^*\ldots x_n^*}$,
  corresponds to the maximum-weight independent set, namely $\wmis(G) =
  \{i: x_i^* = 0\}$\footnote{Notice we use $x_i=\frac{1+s_i}{2}$ instead of $x_i=\frac{1-s_i}{2}$.}.

\section{\desev{} and CK Graphs}
\label{sec:desev}
In this section, we describe a visualization tool, called Decomposed State Evolution Visualization (\desev), which aims to 
``open up''  the quantum evolution black-box from a computational point of view. 
Consider the above adiabatic algorithm for MIS. 
Recall that according to the adiabatic theorem, if the evolution is slow enough, the system remains in the instantaneous ground state.
Let $\ket{\psi(s)}$ be the ground state of $\ham(s)$, for $s \in [0,1]$. For a system of $n$-qubits, $\ket{\psi(s)}$ is a superposition
of $2^n$ possible computational states, namely,
$$\ket{\psi(s)} = \sum_{x \in \{0,1\}^n} \alpha_x(s)
\ket{x}, \quad \quad \text{ where }\sum_{x \in
  \{0,1\}^n}|\alpha_x(s)|^2 =1.$$
For example, we have the initial ground state $\ket{\psi(0)} = \frac{1}{\sqrt{2^n}} \sum_{x \in \{0,1\}^n}\ket{x}$, which is uniform superposition of all $2^n$ states, while the final ground state $\ket{\psi(1)}= \ket{x_1^*x_2^*\ldots x_n^*}$, corresponding to the solution state.
A natural question is: what are the instantaneous ground states $\ket{\psi(s)}$, for $0<s<1$, like? In particular, we would like to ``see''
how the instantaneous ground state evolves? 
A naive solution would be to trace the $2^n$ amplitudes $\alpha_x$.
The task becomes unmanageable even for $n=10$, which has $1024$ amplitudes, even thouhg many may be negligible (close to zero). 

To make the ``visualization''  feasible, we introduce a new measure $\Gamma_k$.
Suppose that $\ham(1)$, has $(m+1) \le 2^n$
distinct energy levels: $E_0 < E_1 < \ldots < E_m$.
For $0 \le k \le m$, 
let $D_k= \{x \in \{0,1\}^n: \ham(1)\ket{x} = E_k \ket{x}\}$ be the set of (degenerate)  computational states that have the same energy level $E_k$ (with respect to  the problem Hamiltonian $\ham(1)$), and
define 
$$ \Gamma_k(s) = \sum_{x \in D_k} |\alpha_x(s)|^2.$$
In other words, $\Gamma_k(s)$ is the total percentage of (computational) 
states of the same energy level $E_k$ participating in $\ket{\psi(s)}$. 
 The idea is now to trace $\Gamma_k$  instead of $\alpha_x$. 
Here we remark that  $\Gamma_k$ are defined for any eigenstate $\ket{\psi}$ and not just for the ground state.


For our purpose, 
we constructed a special
family of vertex-weighted graphs for the MIS problem, called CK graphs~\cite{CK08}. 
We designed the problem instances such that the global minimum is ``hidden'' in the sense that 
 there are many local minima to mislead local search based algorithms. 
Note that  the size of th smallest  instances needs to be necessarily
smaller than $20$ as we are relying on the eigenvalue computation (or
exact diagonization) to
compute $\Gamma_k$. 

\begin{figure}[h]
$$
  \begin{array}{cc}
\includegraphics[width=0.5\textwidth]{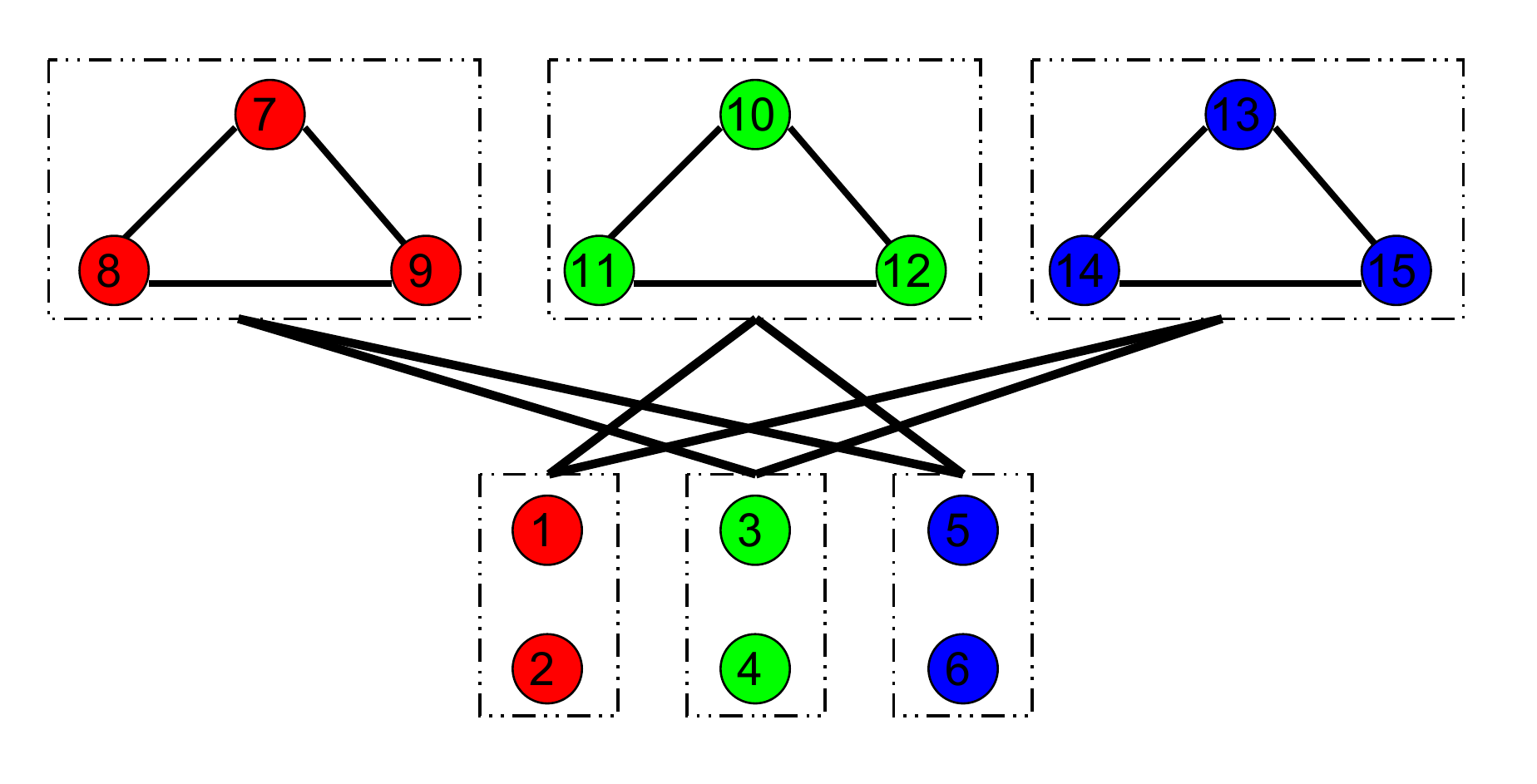} &
\includegraphics[width=0.4\textwidth]{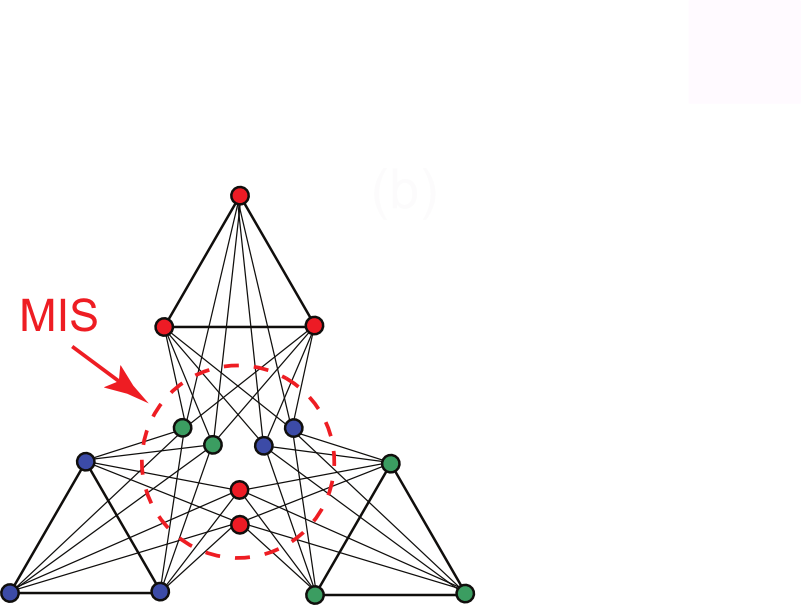}    \\
(a) & (b)
  \end{array}
$$
  \caption{(a) A CK graph for $r=3$ and $g=3$. The graph consists
  of 15 vertices:  $V_A=\{1,\ldots, 6\}$  forms an
    independent set of size 6, while $V_B$, consisting of $g(=3)$ groups of $r(=3)$ triangles:
    $\{7,8,9\}$, $\{10,11,12\}$, and $\{13,14,15\}$, forms $3^3$
    independent sets of size 3. The graph is connected as follows. The 6
    vertices in $V_A$ are divided into 3 groups: $\{1,2\}$, $\{3,4\}$, and
    $\{5,6\}$. The vertices in each group are adjacent to vertices in
    two groups of three triangles in $V_B$ (as illustrated by
  different colors). 
(b) The drawing of the graph with explicit connections.
The weight of a vertex in $V_A$ ($V_B$ resp.) is $w_A$ ($w_B$ resp.).
We set $w_A=1$,  and consider $1 \le w_B < 2$.
For explanation purpose, we represent a vertex in $V_A$ by a $\bullet$, and a vertex in $V_B$ by a $\triangle$.
Therefore, $V_A=\{\bullet,\bullet,\bullet,\bullet,\bullet,\bullet,\}$, forms the MIS of weight $6$; while
$\{\triangle, \triangle, \triangle\}$ is a maximal independent set of weight $3w_B(<6)$. 
  }
\label{fig:G1}
\end{figure}

\noindent{\bf CK Graph Construction.}
Let $r, g$ be integers, and $w_A$, $w_B$ be positive rational
numbers. Our graphs are specified by these four parameters. There
are two types of vertices in the graph: vertices of a $2g$-independent
set, denoted by $V_A$, and vertices of $g$ $r$-cliques
(which form $r^g$ maximal independent sets), denoted by $V_B$.
The weight of 
vertex in $V_A$ ($V_B$ resp.) is $w_A$ ($w_B$ resp.). The connections
between $V_A$ and $V_B$: partition the $2g$ vertices in $V_A$ into
$g$ groups of $2$. Label the $g$ $r$-cliques such that each group is
adjacent to all but one (the same label) $r$-cliques. 
Note if $w_B<2w_A$,
then we have $V_A$ forming the (global) maximum independent sets of
weight $2gw_A$, while there are $r^g$ (local) maximal independent
set of weight $gw_B$. 
See Figure
\ref{fig:G1} for an example of a graph  for
$r=3$ and $g=3$.

\subsection{\desev{} for the MIS{} Adiabatic Algorithm on  a 15-vertex CK Graph}
In the section, we fix the CK graph with $r=3$, $g=3$ as illustrated in Figure~\ref{fig:G1}.
We set $w_A=1$,  and consider $1 \le w_B < 2$.
The graph $\GCK$ consists
  of 15 vertices:  $V_A=\{1,\ldots, 6\}$  forms the maximum-weight
    independent set of weight $6$; 
while $V_B$, consisting of $3$ groups of $3$ triangles:
    $\{7,8,9\}$, $\{10,11,12\}$, and $\{13,14,15\}$, forms  $3^3$ maximal 
    independent sets of weight $3w_B < 6$.

According to Eq.\eqref{eq:Ising}, the problem Hamiltonian (and thus the adiabatic algorithm)
for MIS on $\GCK$ is 
\begin{equation}
\ham_{1} = \sum_{i \in V_A} (6J -2) \sigma_i^z + \sum_{i \in V_B} (6J -2w_B) \sigma_i^z 
+ J\sum_{ij \in \edge(G)} \sigma^z_i \sigma^z_j
\label{eq:unscaled}
\end{equation}
Here we fix $J_{ij}=J=2>w_B$ for all $ij \in \edge(\GCK)$.

\paragraph{Notation on represent the computational states.} For  a computational state $\ket{x_1x_2 \ldots x_n}$ where $x_i \in \{0,1\}$,
we adopt the zero position representation, namely, represent it by $\ket{i_1i_2\ldots i_k}$ where $x_{j}=0$ if and only if $j=i_t$ for some $t$.
That is, we  represent 
$\ket{000000111111111}$ (the solution state) by  $\ket{123456}$. 
Further,  we  use a $\bullet$ to denote a vertex in $V_A$, a $\triangle$ for
a vertex in $V_B$. 
That is, the solution state is now represented
by $\ket{\bullet \bullet \bullet \bullet \bullet \bullet}$,
while $\ket{\triangle \triangle \triangle}$, corresponding to a local maximal independent set of weight $3w_B$ with one vertex from each triangle.

\paragraph{Maximum vs Minimum.} The maximum of MIS corresponds to the minimum of the Ising energy. 
For explanation purpose, instead of referring to the  energy values of the Ising Hamiltonian, we will refer  to
 the values of MIS given by the pseudo-boolean function $\oy$ in Eq.\eqref{eq:Y} by ``(-)energy'', where ``(-)'' is to    indicate the reverse ordering.

\paragraph{Example.} The (-)energy of $\ket{\bullet \bullet \bullet \bullet \bullet \bullet}$ is 6; while $\ket{\triangle \triangle \triangle\!\!\!-\!\!\!\triangle}$ is $4w_B - J$, where $\triangle\!\!\!-\!\!\!\triangle$ represents two connected vertices from $V_B$, e.g. vertex 7 and 8 in Figure~\ref{fig:G1}.

\begin{figure}[h]
\begin{center}
 \includegraphics[width=0.8\textwidth]{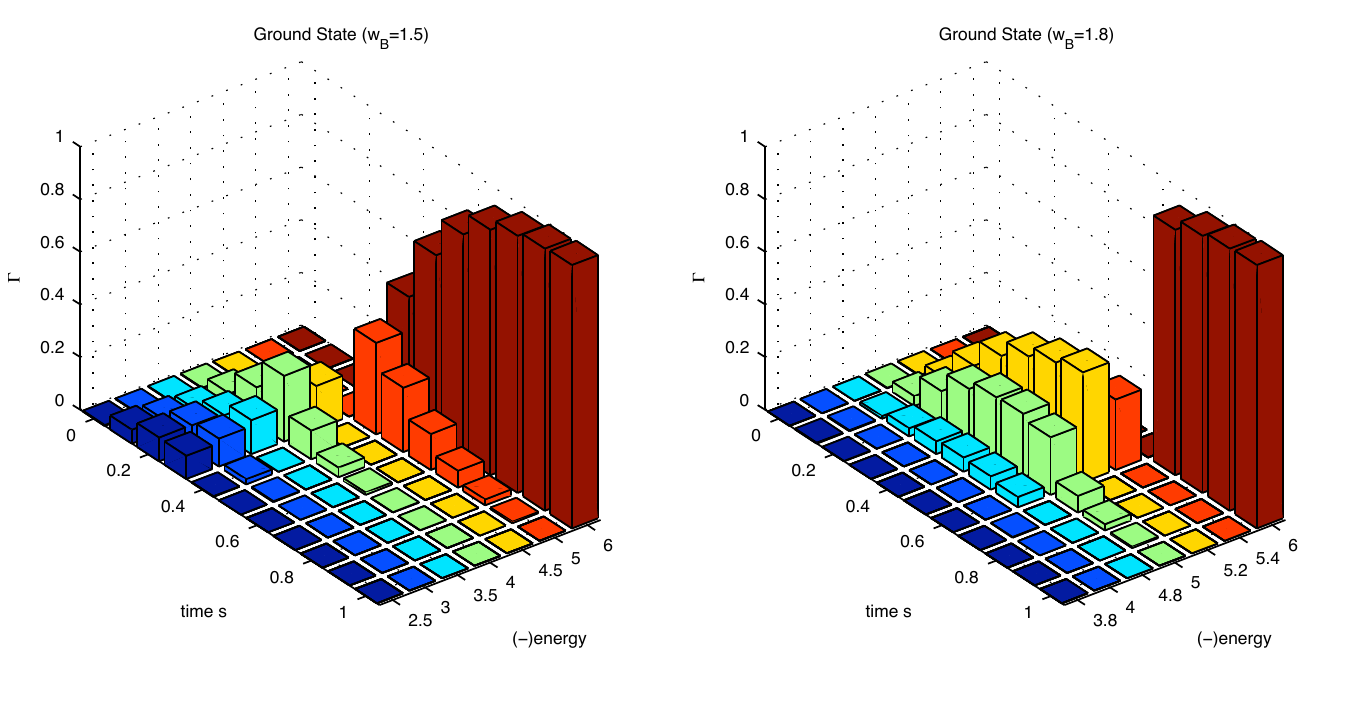}
 $s^*=0.3805, \gmin = 2.04\times10^{-2}$ \hspace*{1cm} $s^*=0.6276, \gmin = 1.04\times10^{-5}$ 
\begin{tabular}{cl||cl}
(-)energy & state & (-)energy & state\\
6 & $\ket{\bullet \bullet \bullet \bullet \bullet \bullet}$ &6 & $\ket{\bullet \bullet \bullet \bullet \bullet \bullet}$\\
5 & $\ket{\bullet \bullet \bullet \bullet \bullet}$ & 5.4 & $\ket{\triangle \triangle \triangle}$\\
4.5 & $\ket{\triangle \triangle \triangle}$ & 5.2 & $\ket{\triangle \triangle \triangle\!\!\!-\!\!\!\triangle}$\\
4 & $\ket{\bullet \bullet \bullet \bullet}$ & 5 & $\ket{\bullet \bullet \bullet \bullet \bullet}$ +  $\ket{\triangle \triangle\!\!\!-\!\!\!\triangle \triangle\!\!\!-\!\!\!\triangle}$\\
3.5 & $\ket{\bullet \bullet \triangle}$ & 4.8 &$\ket{\triangle\!\!\!-\!\!\!\triangle \triangle\!\!\!-\!\!\!\triangle \triangle\!\!\!-\!\!\!\triangle}$\\
3 & $\ket{\bullet \bullet \bullet}$ & 4 & $\ket{\bullet \bullet \bullet \bullet}$\\
2.5 & $\ket{\bullet \triangle}$ & 3.8  & $\ket{\bullet \bullet \triangle}$
\end{tabular}

\end{center}
 \caption{\desev{} (only the 7 lowest energy levels shown) of the ground state of the MIS adiabatic algorithm with $\ham_{1}$ in Eq.\eqref{eq:unscaled} as the problem Hamiltonian for
 $w_B=1.5$ (left) and $w_B=1.8$ (right).
The x-axis is the time $s$. The y-axis is the (-)energy level.
 Each color corresponds to an energy level. 
The correspondence between (-)energy levels and the states are shown.
The z-axis is $\Gamma$.
As time $s$ increases, one can see how $\Gamma$ of each energy level  evolves to get some sense of the evolution.
For example, for $w_B=1.5$ (left), 
for the (-)energy level 6 (which corresponds to the solution state), shown in brown, $\Gamma$ changes from almost 0 at $s=0.2$, to more than $0.4$ at $s=0.4$, 
to almost $1.0$ at $s=0.8$.
For $w_B=1.8$ (right), $\Gamma$ of (-) energy level 6 changes from almost $0$ before $s=0.6$ to more than $0.9$ at $s=0.7$;
while
$\Gamma$ of (-) energy level 5.4, which corresponds to the local minima, gradually increases from $s=0$ to $0.6$,
but almost $0$ after $s=0.6$.
}
  \label{fig:gsd}
\end{figure}

See Figure~\ref{fig:gsd} for the \desev{} of the 
the ground state of the adiabatic algorithm with $\ham_{1}$ in Eq.\eqref{eq:unscaled} as the problem Hamiltonian
for $w_B=1.5$ and $1.8$.


\subsection{FQPT and Perturbation Estimation}
To gain better understanding, we vary the weights of vertices: fix
$w_A = 1$, while varying $w_B$ from $1$ to $1.9$ with a step size of $0.1$. That is, we
fix the global maximum independent set, while increasing the weight of
the local
maximum. 
As the weight of $w_B$ increases, 
the minimum spectral gaps get smaller and
smaller (indeed, from $10^{-1}$ to $10^{-8}$ as $w_B$ changes from $1$ to
$1.9$). See Table~\ref{table1} in Appendix A.

  \begin{figure}
$$
\begin{array}{ccl}
& (\mbox{Zoom:} s=0.627 \dots 0.628) &\\
\includegraphics[width=0.33\textwidth]{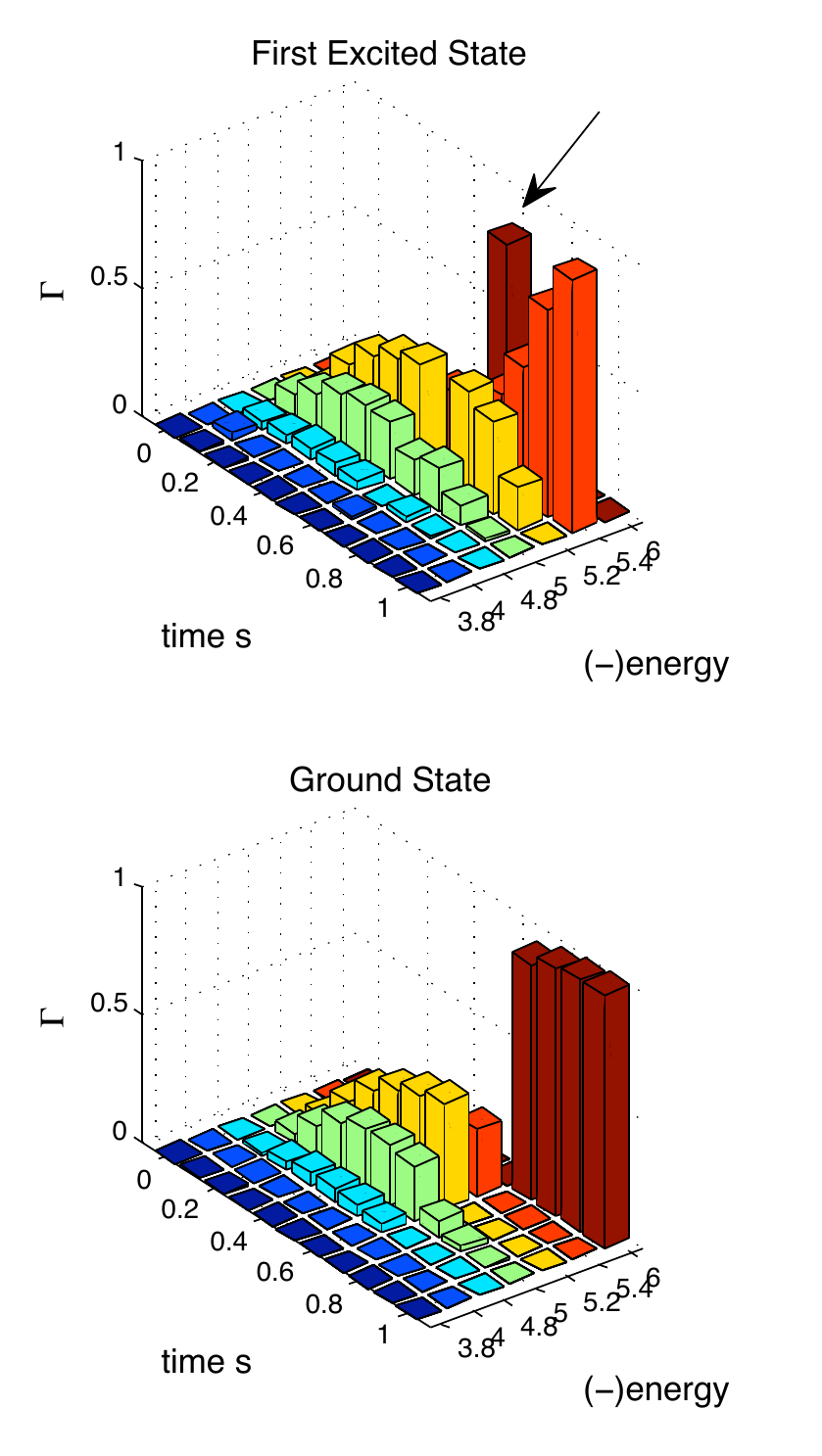}                
&
\includegraphics[width=0.3\textwidth]{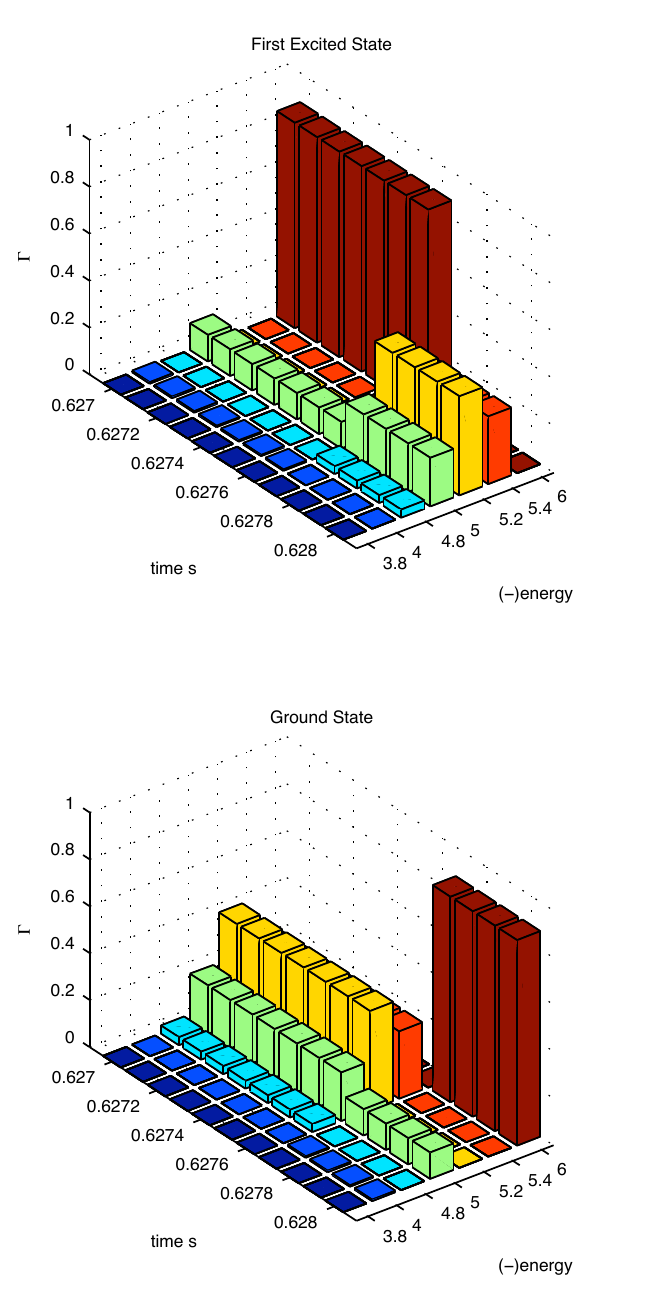}                
&\includegraphics[width=0.3\textwidth]{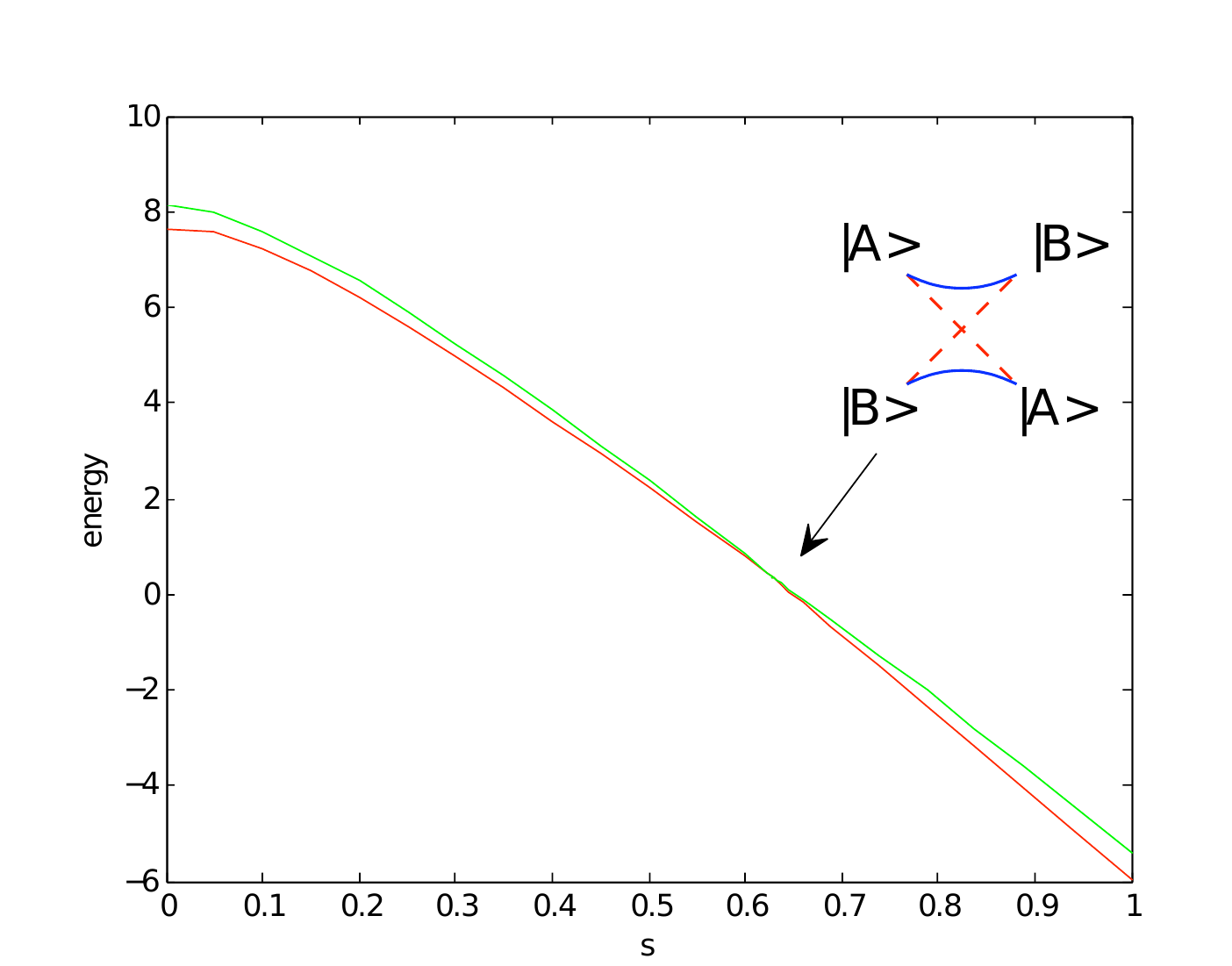}   \\
s^*=0.6276, \gmin = 1.04\times10^{-5}           & &\\
(a) & (b) & (c)
\end{array}
$$
{\tiny
  \begin{tabular}{llllllll}
     & 3.8 & 4 & 4.8 & 5 & 5.2 & 5.4& 6 \\
& $\ket{\bullet \bullet \triangle}$ & $\ket{\bullet \bullet \bullet \bullet}$ & $\ket{\triangle\!\!\!-\!\!\!\triangle \triangle\!\!\!-\!\!\!\triangle \triangle\!\!\!-\!\!\!\triangle}$ & $\ket{\bullet \bullet \bullet \bullet \bullet}$ +  $\ket{\triangle \triangle\!\!\!-\!\!\!\triangle \triangle\!\!\!-\!\!\!\triangle}$ & $\ket{\triangle \triangle \triangle\!\!\!-\!\!\!\triangle}$ & $\ket{\triangle \triangle \triangle}$ &
 $\ket{\bullet \bullet \bullet \bullet \bullet \bullet}$\\ 
 \end{tabular}
}

  \caption{\desev{} of the ground state and the first excited state of the MIS adiabatic algorithm with $\ham_{1}$ in \eqref{eq:unscaled} as the problem Hamiltonian
for $w_B=1.8$  (a) $s=0 \ldots 1$; (b) Zoom in $s=0.627 \ldots 0.628$; (c) The lowest two energy levels of $\ham(s)$, $s=0 \ldots 1$. 
The inset illustrates a level anti-crossing between two states $\ket{B}$ and $\ket{A}$, or the system has a FQPT from $\ket{B}$ to $\ket{A}$ at the anti-crossing $s^*$. 
In this example,
$\ket{A} = \ket{\bullet \bullet \bullet \bullet \bullet \bullet}
+ \ket{\bullet \bullet \bullet \bullet \bullet}$ and $\ket{B} = \ket{\triangle \triangle\!\!\!-\!\!\!\triangle \triangle\!\!\!-\!\!\!\triangle} + \ket{\triangle \triangle \triangle\!\!\!-\!\!\!\triangle} + \ket{\triangle \triangle \triangle}$.
}
  \label{fig:desev1}
\end{figure}

This was consequently explained by the FQPT in \cite{AC09}. 
By FQPT,  here we mean that there is a level anti-crossing between two states as illustrated in 
Figure~\ref{fig:desev1}.
The minimum spectral gap ($\gmin$) and the position ($s^*$)
were then estimated  based on the assumption of the level anti-crossing between the global minimum
and the local minima using perturbation method.
In particular, $\gmin$ was estimated by the tunneling amplitude between the global minimum and the local minima.
The formula so derived involves combinatorial enumeration of the all possible paths between local
minima  and the global minimum, and suggested $\gmin$ is exponentially (in terms of the problem size) small.
 See also \cite{altshuler-2009,AC09,farhi-2009, young-2009} for more explanation on the FQPT and the level anti-crossing.

\section{Varying Parameters in the Problem Hamiltonian for MIS}
\label{sec:change-parameter}
In this section, we show that by changing the parameters in the problem Hamiltonian for MIS on CK graphs, 
the FQPT no longer occurs and 
we can significantly increase $\gmin$.



Recall that in the pseudo-boolean formulation of MIS as in Theorem~\ref{thm:mis}, 
 the requirement for $J_{ij}$ is at least $\min\{c_i,c_j\}$, for each $ij \in \edge(G)$.
For simplicity,  we consider the simplest case in which $J_{ij}=J$ for all $ij \in \edge(G)$.
In other words, we have the corresponding problem Hamiltonian:
\begin{equation*}
\ham_{1} = \sum_{i \in \ver(G)} (d_iJ-2c_i) \sigma^z_i + \sum_{ij \in \edge(G)} J
\sigma^z_i \sigma^z_j\\
\end{equation*}
where $d_i$ is the degree of vertex $i \in \ver(G)$.

The natural question is how does the \art{} change when we vary $J$? Note that it is not sufficient to
consider only the minimum spectral gap change (as almost all the other works on adiabatic quantum computation did)
because by increasing $J$, the maximum energy of the system Hamiltonian also increases.
Instead, in order to keep the maximum energy of the system Hamiltonian comparable, we keep $J$ fixed and
vary $c_i$  instead, namely multiplying all weights $c_i$ by a scaling factor, say  $1/k$, for $k \ge 1$, 
which does not change the original
problem to be solved. We remark that this is equivalent to multiplying $J$ by $k$, and then multiply the problem Hamiltonian by ($1/k$). The details and more general case can be found in \cite{Precision-scale}.

That is, we consider the following 
(scaled) problem Hamiltonian 
\begin{eqnarray}
\ham_k= \sum_{i \in \ver(G)} (Jd_i-2c_i/k )\sigma^z_i + \sum_{ij \in \edge(G)} J 
\sigma^z_i \sigma^z_j
\label{eq:Hk}
\end{eqnarray}
where $k \ge 1$ is the scaling factor.

\subsection{Minimum Spectral Gap $\gmin$ Without FQPT}
\begin{figure}[h]
$$
\begin{array}{cc}
k=1 & k=10 \\
\includegraphics[width=0.45\textwidth]{K1_annotate.pdf}                
&
\includegraphics[width=0.4\textwidth]{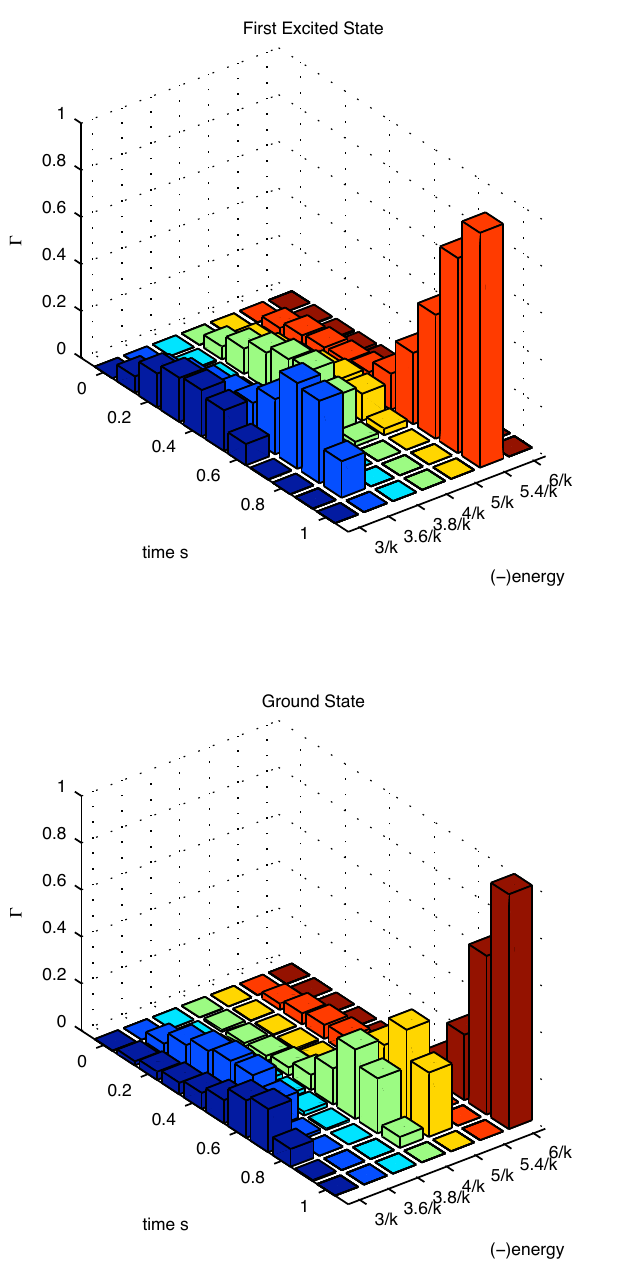} \\
s^* = 0.627637, \gmin=1.04\times10^{-5} & s^*=0.667731, \gmin=0.145
\end{array}
$$

{\tiny
  \begin{tabular}{llllllll}
    $k=1$ & 3.8 & 4 & 4.8 & 5 & 5.2 & 5.4& 6 \\
& $\ket{\bullet \bullet \triangle}$ & $\ket{\bullet \bullet \bullet \bullet}$ & $\ket{\triangle\!\!\!-\!\!\!\triangle \triangle\!\!\!-\!\!\!\triangle \triangle\!\!\!-\!\!\!\triangle}$ & $\ket{\bullet \bullet \bullet \bullet \bullet}$ +  $\ket{\triangle \triangle\!\!\!-\!\!\!\triangle \triangle\!\!\!-\!\!\!\triangle}$ & $\ket{\triangle \triangle \triangle\!\!\!-\!\!\!\triangle}$ & $\ket{\triangle \triangle \triangle}$ &
 $\ket{\bullet \bullet \bullet \bullet \bullet \bullet}$\\ \\
$k=10$ & $3/k$ & $3.6/k$ & $3.8/k$ & $4/k$ & $5/k$ & $5.4/k$ & $6/k$\\
& $\ket{\bullet \bullet \bullet}$ & $\ket{\triangle \triangle}$ & $\ket{\bullet \bullet \triangle}$ &
$\ket{\bullet \bullet \bullet \bullet}$ & $\ket{\bullet \bullet \bullet \bullet \bullet}$ & $\ket{\triangle \triangle \triangle}$
& $\ket{\bullet \bullet \bullet \bullet \bullet \bullet}$
 \end{tabular}
}
\caption{\desev{} of the ground state and the first excited state of the MIS adiabatic algorithm with  problem Hamiltonian $\ham_{1}$ (left)
and  $\ham_{10}$ (right)
where $w_B=1.8$. 
Notice the differences in the lowest few excited states.
For $k=1$, the $2nd$ and $3rd$ excited states are  superpositions of $\triangle$s (vertex in $V_B$ which constitutes the local optima);
while for $k=10$, the $2nd$ and $3rd$ excited states are  superpositions of $\bullet$s (vertex in $V_A$ which constitutes the global optimum). As a result, the first order phase transition from local minima to global minimum occurs for $k=1$, which results in the $\gmin= 1.04\times 10^{-5}$ at $s^*=0.627$. For $k=10$, such crossing no longer occurs, and $\gmin=0.145$ at $s^*=0.667$.  
See Figure~\ref{fig:scaled-zoom} for the zoom-in.
}
  \label{fig:scaled}
\end{figure}

\begin{figure}[h]
$$
\begin{array}{ccc}
k=1 (s:0.627 \ldots 0.628)  & k=1 (s:0.62763 \ldots 0.62764)  &k=10 (s: 0.667 \ldots 0.668)\\
\includegraphics[width=0.3\textwidth]{K1_Zoom.pdf}                
&
\includegraphics[width=0.3\textwidth]{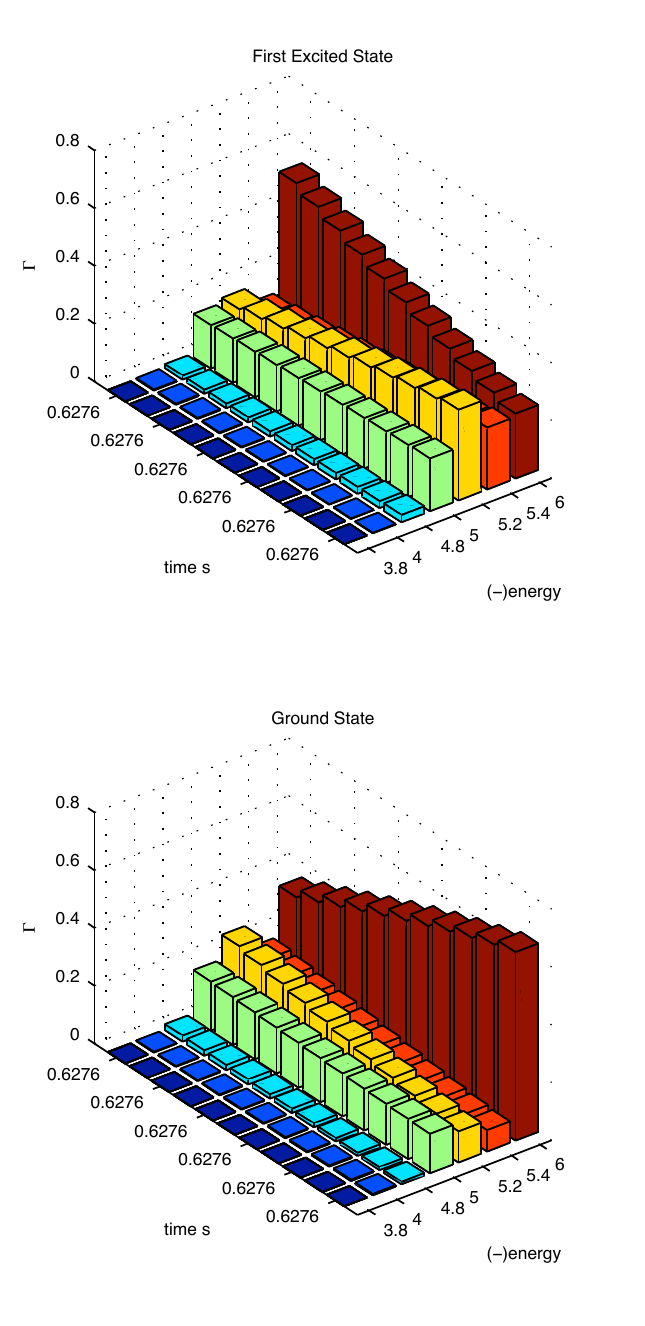}                
&
\includegraphics[width=0.3\textwidth]{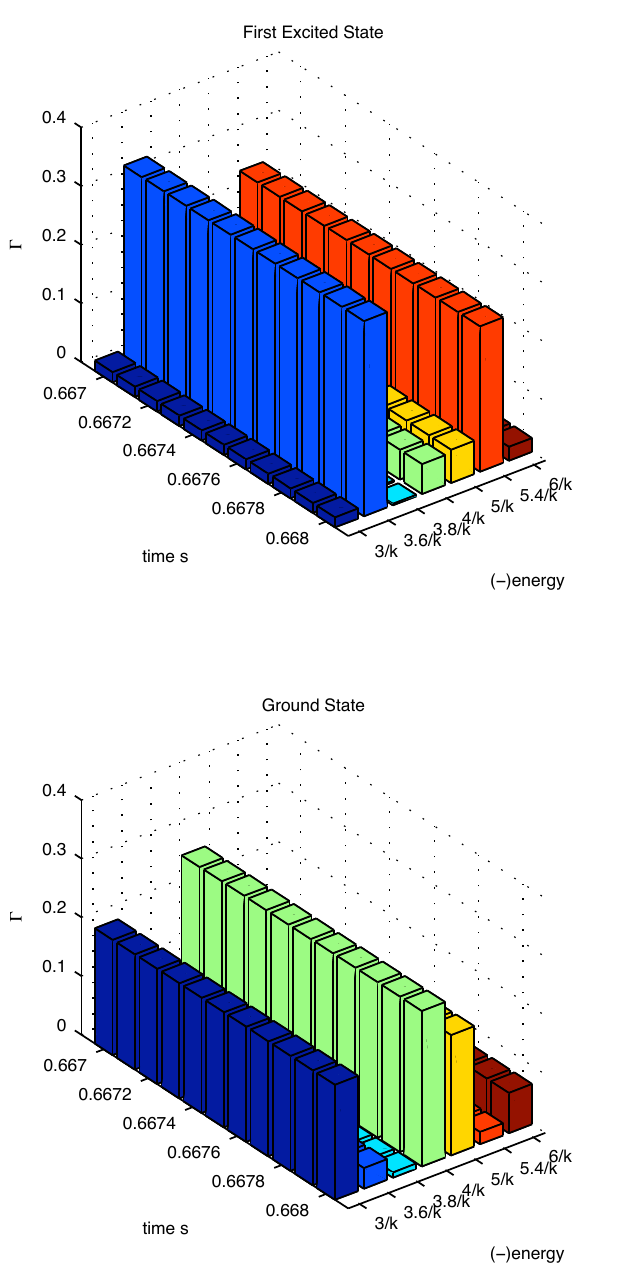}                \\
s^*= 0.62763727, \gmin=1.04\times 10^{-5} & s^*= 0.62763727, \gmin=1.04\times 10^{-5}  & s^*=0.66773072, \gmin=1.45 \times 10^{-1}\\

\end{array}
 $$
{\tiny
  \begin{tabular}{llllllll}
    $k=1$ & 3.8 & 4 & 4.8 & 5 & 5.2 & 5.4& 6 \\
& $\ket{\bullet \bullet \triangle}$ & $\ket{\bullet \bullet \bullet \bullet}$ & $\ket{\triangle\!\!\!-\!\!\!\triangle \triangle\!\!\!-\!\!\!\triangle \triangle\!\!\!-\!\!\!\triangle}$ & $\ket{\bullet \bullet \bullet \bullet \bullet}$ +  $\ket{\triangle \triangle\!\!\!-\!\!\!\triangle \triangle\!\!\!-\!\!\!\triangle}$ & $\ket{\triangle \triangle \triangle\!\!\!-\!\!\!\triangle}$ & $\ket{\triangle \triangle \triangle}$ &
 $\ket{\bullet \bullet \bullet \bullet \bullet \bullet}$\\ \\
$k=10 $ & $3/k$ & $3.6/k$ & $3.8/k$ & $4/k$ & $5/k$ & $5.4/k$ & $6/k$\\
& $\ket{\bullet \bullet \bullet}$ & $\ket{\triangle \triangle}$ & $\ket{\bullet \bullet \triangle}$ &
$\ket{\bullet \bullet \bullet \bullet}$ & $\ket{\bullet \bullet \bullet \bullet \bullet}$ & $\ket{\triangle \triangle \triangle}$
& $\ket{\bullet \bullet \bullet \bullet \bullet \bullet}$
 \end{tabular}
}

  \caption{Zoom around $s^*$. }
  \label{fig:scaled-zoom}
\end{figure}

\begin{figure}
$$
\begin{array}{ccc}
k=1 & k=2 &k=3\\
\includegraphics[width=0.3\textwidth]{K1_annotate.pdf} & \includegraphics[width=0.25\textwidth]{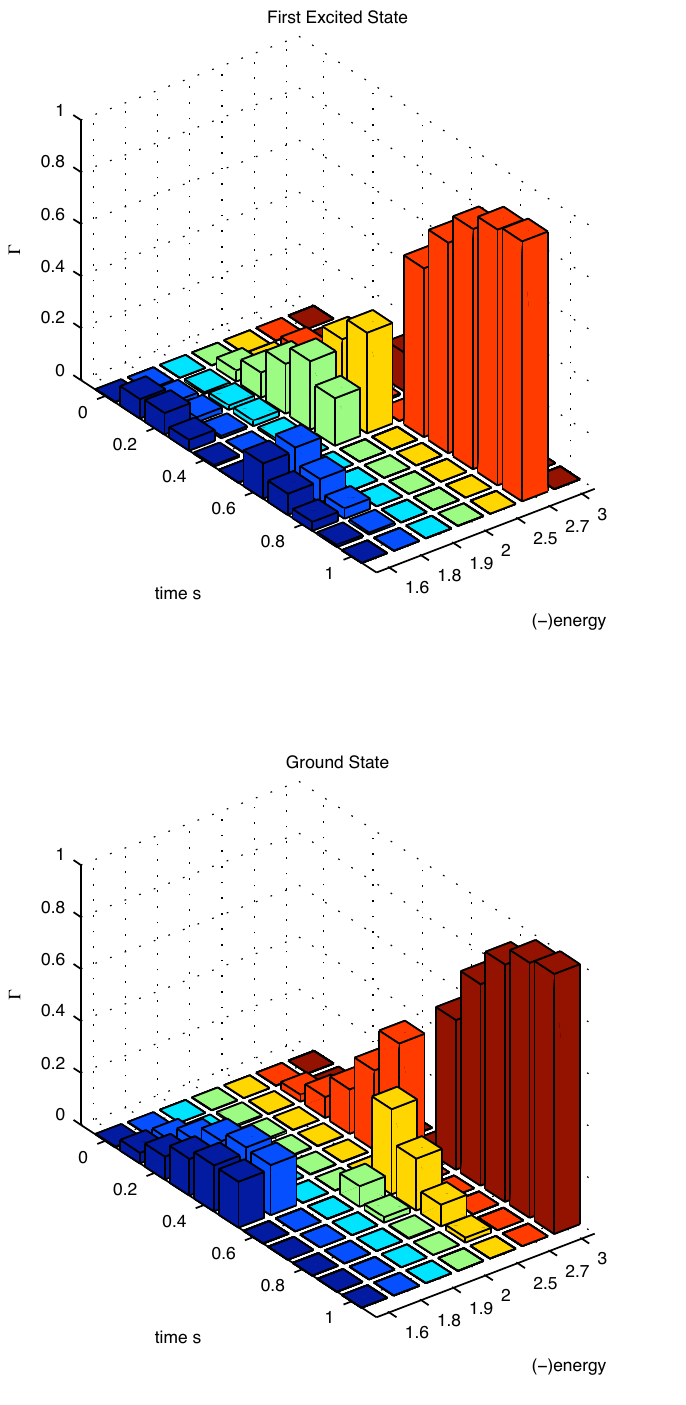} & \includegraphics[width=0.25\textwidth]{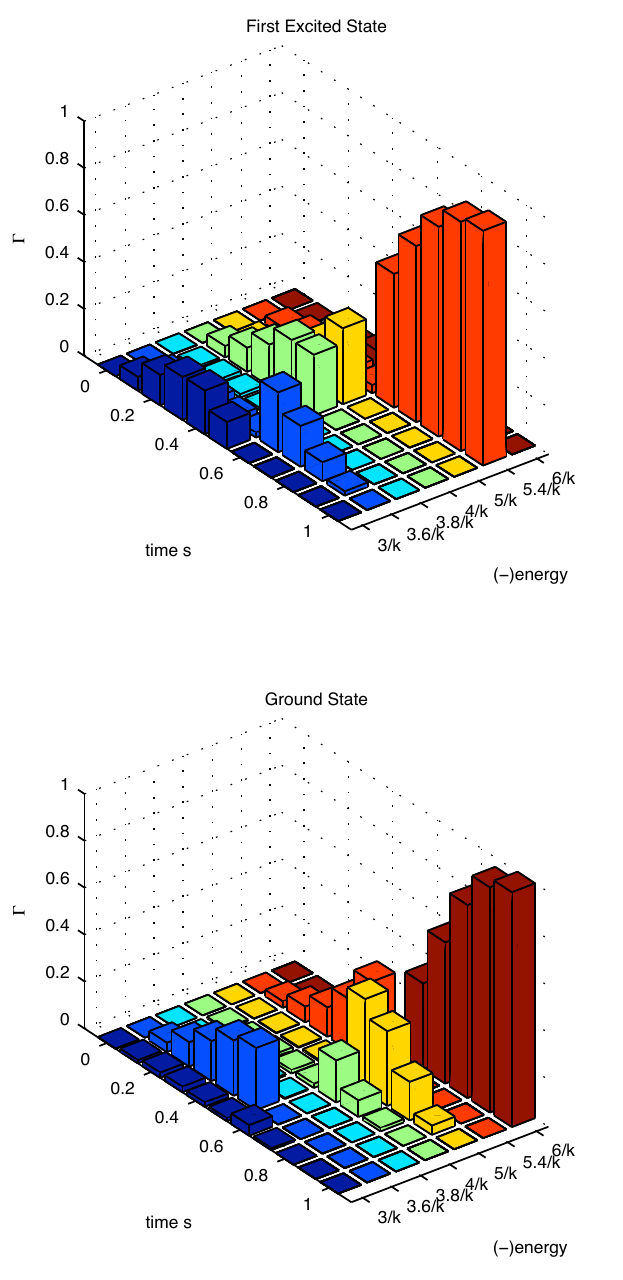}\\
s^*= 0.62763727, \gmin=1.04\times 10^{-5} & s^*=0.54578285,	\gmin=6.37\times 10^{-3}  & s^*=0.54467568,	\gmin=3.30\times10^{-2}\\
k=5 & k=10 & k=50\\
\includegraphics[width=0.25\textwidth]{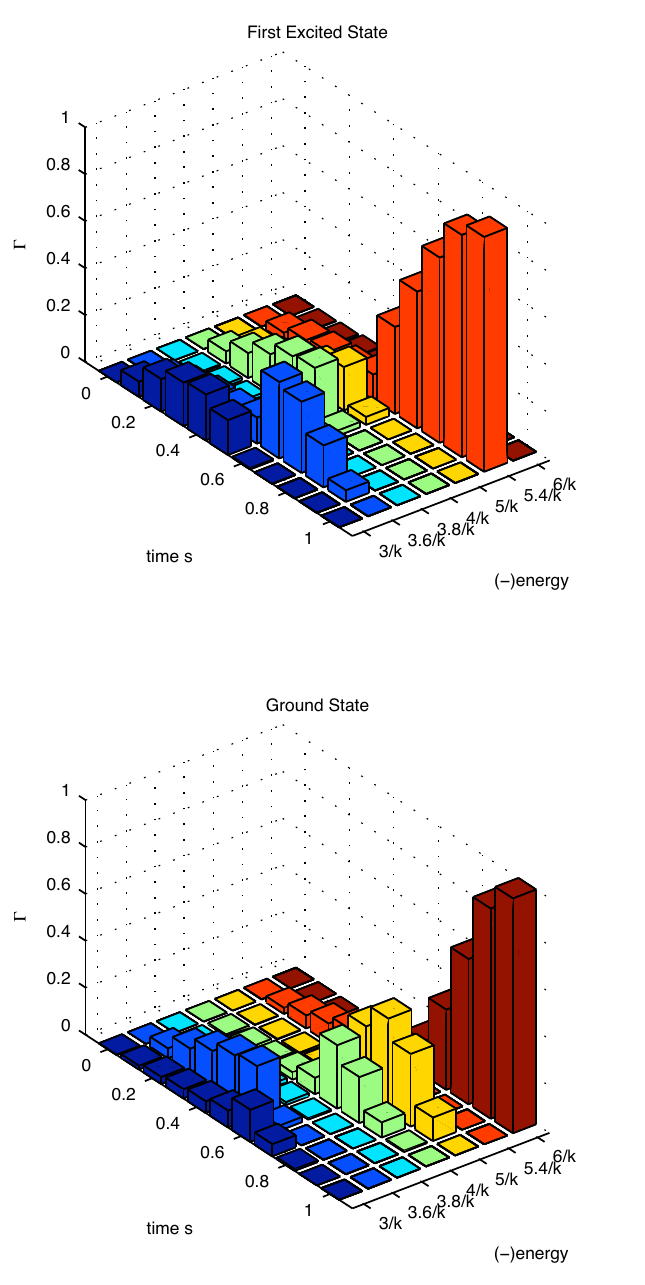} & \includegraphics[width=0.25\textwidth]{K10.pdf} & \includegraphics[width=0.25\textwidth]{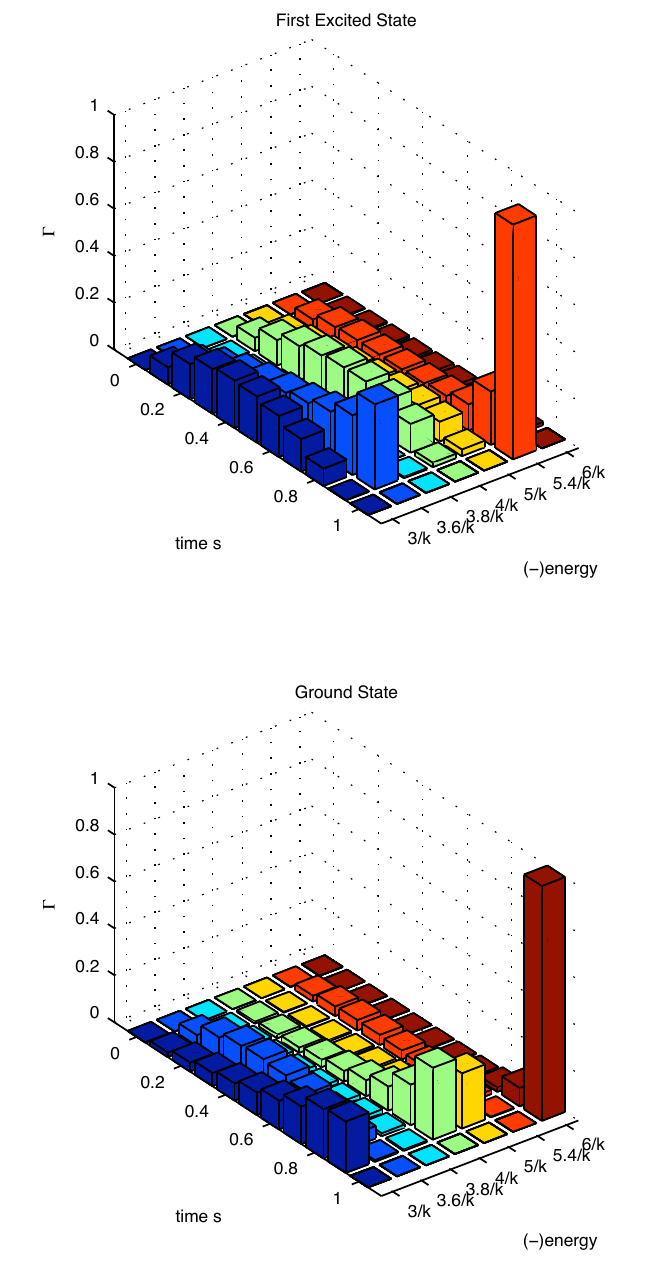}\\
s^*=0.57419149, \gmin=9.67\times 10^{-2} & s^*=0.66773072, \gmin=1.45 \times 10^{-1} & s^*=0.99779592,\gmin=4.79 \times 10^{-2}
\end{array}
$$ 
{\tiny
  \begin{tabular}{llllllll}
    $k=1$ & 3.8 & 4 & 4.8 & 5 & 5.2 & 5.4& 6 \\
& $\ket{\bullet \bullet \triangle}$ & $\ket{\bullet \bullet \bullet \bullet}$ & $\ket{\triangle\!\!\!-\!\!\!\triangle \triangle\!\!\!-\!\!\!\triangle \triangle\!\!\!-\!\!\!\triangle}$ & $\ket{\bullet \bullet \bullet \bullet \bullet}$ +  $\ket{\triangle \triangle\!\!\!-\!\!\!\triangle \triangle\!\!\!-\!\!\!\triangle}$ & $\ket{\triangle \triangle \triangle\!\!\!-\!\!\!\triangle}$ & $\ket{\triangle \triangle \triangle}$ &
 $\ket{\bullet \bullet \bullet \bullet \bullet \bullet}$\\ \\
$k=2 $ & $1.6$ & $1.8$ & $1.9$ & $2$ & $2.5$ & $2.7$ & $3$\\
& $\ket{\triangle \triangle\!\!\!-\!\!\!\triangle}$ & $\ket{\triangle \triangle}$ & $\ket{\bullet \bullet \triangle}$ &
$\ket{\bullet \bullet \bullet \bullet}$ & $\ket{\bullet \bullet \bullet \bullet \bullet}$ & $\ket{\triangle \triangle \triangle}$
& $\ket{\bullet \bullet \bullet \bullet \bullet \bullet}$\\ \\

$k\ge3 $ & $3/k$ & $3.6/k$ & $3.8/k$ & $4/k$ & $5/k$ & $5.4/k$ & $6/k$\\
& $\ket{\bullet \bullet \bullet}$ & $\ket{\triangle \triangle}$ & $\ket{\bullet \bullet \triangle}$ &
$\ket{\bullet \bullet \bullet \bullet}$ & $\ket{\bullet \bullet \bullet \bullet \bullet}$ & $\ket{\triangle \triangle \triangle}$
& $\ket{\bullet \bullet \bullet \bullet \bullet \bullet}$
 \end{tabular}
}

 \caption{\desev{} of the ground state and the first excited state of the adiabatic algorithm with  problem Hamiltonian
$\ham_k$ for $w_B=1.8$, where $k=1,2,3,5,10,50$.}
  \label{fig:different-K}
\end{figure}

The \desev{} of $\ham_{1}$ and $\ham_{10}$ ($k=10$) is shown in Figure~\ref{fig:scaled} and Figure~\ref{fig:scaled-zoom}.
The anti-crossing between the global minimum and the local minima (for $k=1$) 
no longer occurs for $k=10$, and $\gmin$ increases from $1.04\times10^{-5}$ to $0.145$. 
Notice that the change in the lowest few excited energy levels: 
 for $k=1$, the lowest few excited states (beyond the first excited state) 
of the problem Hamiltonian  is mainly the superposition of states from $V_B$ ($\triangle$) (which constitutes the local minima); while 
these states of the scaled ($k=10$) problem Hamiltonian is mainly the superposition of states from $V_A$($\bullet$) (which constitutes the global minimum).
The \desev{} of $\ham_k$ for $k=1,2,3,5,10,50$ is shown in Figure~\ref{fig:different-K}.

In \cite{AC09}, 
based on the FQPT assumption, 
we estimate $\gmin$ (for $\ham_{1}$) by the tunneling amplitude between the local minima and the global minimum,
which suggests that $\gmin$ is exponentially small. 
However, for $k=10$,
from our numerical data and \desev{} in Figure~\ref{fig:scaled}, we see that the FQPT 
(that causes $\gmin$ to be exponentially small)  
no longer 
occurs, and $\gmin$ increases significantly (from $10^{-5}$ to $0.145$).
This seems to suggest that $\gmin$ to be polynomially small instead.
 We are currently investigating how to analytically bound or estimate $\gmin$
 of $\ham_{k}$ for a general CK graph of size $n$.
We remark here that the 
perturbation method is still valid (in fact, 
 as we increase $k$, we also increase the minimum spectral gap position $s^* \leadsto 1$ (see \cite{Precision-scale} for the proof)), however we can no longer assume that $\gmin$ can be approximated by the tunneling amplitude between the two (localized) states.

\subsection{Scaling Factor and \art{}}
\label{sec:art}
In this section, we discuss what  the good scaling factor should be, and how it affects the \art.
To address this question, we need an appropriate formulation for \art.
We point out that it is not sufficient to just consider $\gmin$, but the matrix element of the time 
derivative of the Hamiltonian also matters.
In particular, we adopt the following three formulations,
which are related to the widely used traditional condition:
$$
(*)
\left\{
\begin{array}{l}
  \art_1(\ham) = \frac{\max_{0 \le s \le 1}\mat(s)}{\gmin^2}\max_{0 \le s \le 1}||\ham||  \\
  \art_2(\ham) = \frac{\mat(s^*)}{\gmin^2}\max_{0 \le s \le 1}||\ham||, \mbox{ where } \gmin=E_1(s^*) - E_0(s^*)\\
  \art_3(\ham) = \max_{0 \le s \le 1}\frac{\mat(s)}{(E_1(s) - E_0(s))^2}\max_{0 \le s \le 1}||\ham|| 
\end{array}
\right.
$$
where $\mat(s)=|\bra{E_1(s)}\frac{d\ham}{ds}\ket{E_0(s)}|$ is the matrix element of the time derivative Hamiltonian at time $s$,
and $\ham(s) \ket{E_i(s)} = E_i(s)\ket{E_i(s)}$.
See Table~\ref{table2} for the numerical comparisons.

    \begin{table}[h]
	\begin{tabular}{|l|l|l|l|l|l|l|l|}
\hline
$k$ & $s^*$ & $\gmin $ & $\mat(s^*)$ &  $\max_{0 \le s \le 1}\mat(s)$   & $\max_{0 \le s \le 1}||\ham||$ & $\art_2$ & $\art_1$\\
\hline
1  &	0.62763727 &	1.04e-05 &	4.02e+00 &	4.02e+00 &	2.26e+02 &	8.34e+12 &		8.34e+12 \\ \hline 
2  &	0.54578285 &	6.37e-03 &	2.04e+00 &	1.69e+00 &	2.48e+02 &	1.24e+07 &		1.03e+07 \\ \hline 
3  &	0.54467568 &	3.30e-02 &	1.41e+00 &	1.01e+00 &	2.55e+02 &	3.32e+05 &		2.37e+05 \\ \hline 
4  &	0.55610853 &	6.83e-02 &	1.18e+00 &	1.18e+00 &	2.59e+02 &	6.57e+04 &		6.58e+04 \\ \hline 
5  &	0.57419149 &	9.67e-02 &	1.06e+00 &	1.07e+00 &	2.61e+02 &	2.96e+04 &		2.99e+04 \\ \hline 
10 &	0.66773072 &	1.45e-01 &	7.48e-01 &	7.92e-01 &	2.66e+02 &	9.45e+03 &		1.00e+04 \\ \hline 
20 &	0.80170240 &	1.30e-01 &	4.72e-01 &	5.68e-01 &	2.68e+02 &	7.48e+03 &		9.01e+03 \\ \hline 
30 &	0.99318624 &	7.97e-02 &	8.95e-09 &	4.26e-01 &	2.69e+02 &	3.78e-04 &		1.80e+04 \\ \hline 
40 &	0.99642154 &	5.99e-02 &	4.90e-10 &	4.35e-01 &	2.69e+02 &	3.67e-05 &		3.26e+04 \\ \hline 
50 &	0.99779592 &	4.79e-02 &	5.30e-11 &	4.41e-01 &	2.69e+02 &	6.20e-06 &		5.16e+04 \\ \hline 
	\end{tabular}

\vspace*{0.5cm}

\begin{tabular}{|l|l|l|l|l|l|l|}
\hline
$k$ & $s'$ & $g(s') $ & $\mat(s')$ & $\frac{\mat(s')}{g(s')^2}$    & $\max_{0 \le s \le 1}||\ham||$ & $\art_3$ \\
\hline
1 &	0.62763727 &	1.04e-05 &	4.02e+00 &	3.70e+10 &	2.26e+02 &	8.34e+12 \\ \hline 
2 &	0.54578226 &	6.37e-03	&     2.04e+00	&     5.02e+04	&     2.48e+02	&     1.24e+07 \\ \hline
3 &	0.54461081 &	3.30e-02 &	1.41e+00 &	1.30e+03 &	2.55e+02 &	3.32e+05 \\ \hline 
4 &	0.55545411 &	6.83e-02 &	1.18e+00 &	2.54e+02 &	2.59e+02 &	6.57e+04 \\ \hline 
5 &	0.57223394 &	9.68e-02 &	1.07e+00 &	1.14e+02 &	2.61e+02 &	2.97e+04 \\ \hline 
10&	0.65682886 &	1.46e-01 &	7.75e-01 &	3.64e+01 &	2.66e+02 &	9.66e+03 \\ \hline 
20&	0.77115481 &	1.33e-01 &	5.41e-01 &	3.08e+01 &	2.68e+02 &	8.24e+03 \\ \hline 
30&	0.83962780 &	1.08e-01 &	4.43e-01 &	3.82e+01 &	2.69e+02 &	1.02e+04 \\ \hline 
40&	0.88050519 &	8.82e-02 &	3.93e-01 &	5.05e+01 &	2.69e+02 &	1.36e+04 \\ \hline 
50&	0.90581875 &	7.39e-02 &	3.63e-01 &	6.64e+01 &	2.69e+02 &	1.79e+04 \\ \hline 
	\end{tabular}

where $g(s) = E_1(s) - E_0(s)$, and $s' = \argmax_{0 \le s \le 1}\frac{\mat(s)}{g(s)^2}$. 

\caption{$\art_1$, $\art_2$, $\art_3$ for $\ham_k$  in Eq.\eqref{eq:Hk}.
Observations: (1)  $\gmin$ increases as $k$ increases  from 1 to 10, but decreases from $10$ to $50$. 
(2) $\art_1$, $\art_2$, and $\art_3$ are close for $k<5$.
(3) The matrix element $\mat(s^*)$ at the position of minimum spectral gap is extremely small for $k \ge 30$.
(4) For $k>10$, $s^*$ (the position of the minimum spectral gap) 
is different from $s'$, where $s' = \argmax_{0 \le s \le 1}\frac{\mat(s)}{g(s)^2}$.  
Note that $\art_1$, and $\art_3$ are close, in particular, they coincide for small $k<5$.
}
\label{table2}
    \end{table}


From Table\ref{table2}, we see that $\gmin$ increases as $k$ increases  from 1 to 10, however, decreases from $10$ to $50$ (even though it is still much larger than $k=1$).
The latter, perhaps,  can be explained by the following: 
 as $k$ increases, the difference between the low energy levels decreases, and becomes dominate for $k>10$.
We remark that the optimal value for $k$ seems to depend only on  the vertex weights (for which $J$ depends on),
and  independ of the problem size.
By increasing the scaling factor, we also increase the 
 precision (or dynamic range) requirement for representing the parameters ($h_i$ \& $J_{ij}$) in the problem Hamiltonian, 
which is one of the important physical resources.

The three versions of \art{} look similar,  
and indeed they coincide for some Hamiltonians (e.g. for $k=1$). 
However, they can be very different for the large $k$. The main reason is that the matrix element $\mat(s)$ can be extremely small at the minimum spectral gap position $s^*$. For example, for $k=50$, $s^* \leadsto 1$, $\mat(s^*)$ is extremely small. Note one can show that $\mat(s)=|\bra{E_1(s)} \ham_{\ms{init}}
  \ket{E_0(s}|/s$ for $s\in (0,1]$\footnote{$ \mat(s)= |\bra{E_1(s)}\ham(1) - \ham(0)\ket{E_0(s}| = |\bra{E_1(s)}\frac{\ham(s) - \ham(0)}{s}\ket{E_0(s}| =|\bra{E_1(s)}\ham(0)\ket{E_0(s}|/s$. See \cite{Precision-scale} for more details.}.
  Thus, for our initial Hamiltonian, $\mat(s)$ measures the overlap of the states with one single bit flip, and in this case it is extremely small.
Observe that
the position of the minimum spectral gap $s^*$ is not the same as the position $s'$ where $\frac{\mat(s)}{g(s)^2}$ is maximized.
What should be the appropriate formulation of $\art$? Should it be $\art_3$? If so, under what condition, can $\art_1$ be a good approximation to $\art_3$?
and under what condition, can we assume that $\gmin$ is the dominating factor (as have been assumed by all other works)? 

\section{Adiabatic Algorithms for \EC and \SAT Based on the Reduction to MIS}
\label{sec:ec-sat}
In this section, we first recall the Exact Cover problem, and then explain the special case of 
\EC as the positive-1-in-3SAT, also called EC3,  for which the adiabatic algorithm was first proposed for.
The problem Hamiltonian of the proposed adiabatic algorithm for EC3 or the general 3SAT (see, e.g., ~\cite{farhi-2009,DV})
 is based on the cost function which computes the number
of clauses violated. This form (i.e. clause-violation cost function) of problem Hamiltonian has been adopted (there are two slightly different forms) by all the existing work. 
In the following, we describe different problem Hamiltonians for \EC and \SAT that are based on the reduction to MIS. 
We then point out that the argument that  the clause-violation cost function based problem Hamiltonian 
(and thus the adiabatic algorithm) for Exact Cover~\cite{altshuler-2009} has exponential small $\gmin$ (and thus require exponential time)
 does not apply to the MIS reduction based problem Hamiltonian.

\subsection{\EC}
Formally, the \EC is as follows:

\hspace*{0.5cm}{\bf Input:} A set of $m$ elements, $X = \{c_1, c_2, \ldots, c_m\}$;  
a family of $n$ subsets of $X$, $\mathcal{S} = \{S_1, S_2, \ldots, S_n\}$, where $S_i \subset X$

\hspace*{0.5cm}{\bf Question:} Is there a subset $I \subseteq \{1, \ldots, n\}$ 
such that $\cup_{i \in I} S_i = X$, where $S_i \cap S_j = \emptyset$ for $i\neq j \in I$? 
Here $\{S_i: i \in I\}$ is called an {\em exact cover} of $X$.

\paragraph{Example.} $X=\{c_1,c_2,c_3,c_4,c_5\}$, and $\mathcal{S} = \{S_1, S_2, \ldots, S_7\}$, with 
$S_1=\{c_1,c_2,c_4\}$, $S_2=\{c_1,c_2,c_5\}$, $S_3=\{c_1,c_3, c_4\}$,$S_4 = \{c_2, c_3\}$, $S_5=\{c_3\}$, $S_6=\{c_4,c_5\}$, $S_7=\{c_5\}$.
Here $\{S_1, S_5, S_7\}$ is the exact cover of $X$.

In particular, if we further restrict that each element $c_i \in X$ appears exactly in three subsets. The problem is referred as
EC3, which can then be polynomially reducible to the positive 1-in-3SAT problem:

\paragraph{\SEC $\le_{P}$ positive 1-in-3SAT.} Given an instance of \SEC with an $m$-element set $X$ and $n$ subsets $S_1, \ldots, S_n$,
we construct a 3CNF boolean formula $\Psi(x_1, \ldots, x_n) = C_1 \wedge \ldots \wedge C_m$ with $n$ variables and $m$ clauses.
Associate each set $S_i$ with a binary variable $x_i$. 
 For each element $c_i \in X$, let $S_{i_1}$, $S_{i_2}$, $S_{i_3}$ be the three sets that consist of $c_i$.
 Define the corresponding clause $C_i = x_{i_1} \vee x_{i_2} \vee x_{i_3}$.
Then if is clear that there is an exact cover to the original problem if and only if the formula 
$\Psi(x_1, \ldots, x_n) = C_1 \wedge \ldots \wedge C_m$ is satisfiable in that there is exactly one variable in each clause is satisfied.

The cost function 
$$
\eng_\Psi(x_1, \ldots, x_n) = \sum_{i=1}^{m} (x_{i_1} + x_{i_2} + x_{i_3} -1 )^2.
$$
penalizes each violating clause. $\Psi$ is satisfiable if and only if the minimum of $\eng_\Psi$ is $zero$ (i.e. no violation).
The corresponding problem Hamiltonian based on this penalty function as used by  Altshuler~et~al.\cite{altshuler-2009} (and Young~et~al.~\cite{young-2009})  \footnote{The sign of $\sigma_i^z$ term is in opposite because they use $x_i=\frac{1-s_i}{2}$ instead. $I_{ij}$ was called $J_{ij} (= \frac{1}{2}(J_{ij} + J_{ji}))$ in \cite{altshuler-2009}.}
\begin{eqnarray}
  \ham_{AY} = \sum_{i \in \ver(\GEC)}B_i \sigma_i^z +  \sum_{ij \in \edge(\GEC)} I_{ij} \sigma_i^z \sigma_j^z
\end{eqnarray}
where $B_i$ is the number of clauses that contains variable $x_i$,  and $I_{ij}$ 
 is the number of clauses that contains both $x_i$ and $x_j$, and  
$\ver(\GEC) = \{1,  \ldots, n\}$,
 and $\edge(\GEC) = \{ij: x_i \mbox{ and } x_j \mbox{ appear in a clause.}\}$.

Next, we show the polynomial reduction from Exact Cover to MIS:
\paragraph{\EC $\le_{P}$ MIS.} Given an instance of \EC with an $m$-element set $X$ and $n$ subsets $S_1, \ldots, S_n$,
 we construct a graph $G_M$ with $n$ vertices, where vertex $i$ corresponds to the set $S_i$. The weight of vertex $i$ is the number of elements in $S_i$. There is an edge between two vertices if and only if $S_i$ and $S_j$ share a common element. 
Thus, there is an exact cover to the original problem if only if  the weight of $\wmis(G_M)$  is $m$.

For EC3, it is easy to see that $\GEC$ and $G_M$ are exactly the same because  there is one-one corresponding between the variable $x_i$ and the set $S_i$ ($\ver(\GEC)=\ver(G_M)$), and  ``$x_i \mbox{ and } x_j$ appear in a  clause'' is equivalent to 
 ``$S_i$ and $S_j$ share a common element'' ($\edge(\GEC)=\edge(G_M)$).
Based on this reduction, we therefore have the following problem Hamiltonian for the same problem:
\begin{eqnarray}
\ham_C= \sum_{i \in \ver(\GEC)} \left(\sum_{j \in \nbr(i)}J_{ij}-2B_i/k \right)\sigma^z_i + \sum_{ij \in \edge(\GEC)} J_{ij} 
\sigma^z_i \sigma^z_j
\end{eqnarray}
where $J_{ij} > \min\{B_i,B_j\}$, and $k \ge1$.

See Appendix A for an example on how to reduce \SEC (given as a \SAT problem) to a MIS problem.
As pointed out by Young~\cite{young-note}, $\ham_{AY}$ and $\ham_C$ are the same for the special case 
$J_{ij}=2I_{ij}$.
However, our
model is much more general since the $J_{ij}$ can take {\it any} values
provided only  that $J_{ij} > \min\{B_i, B_j\}$.   Recall that $2B_i = \sum_{j \in \nbr(i)} I_{ij}$.
In particular, for some $ij \in \edge(G)$, $J_{ij} > 2 I_{ij}$ (e.g. $I_{23}=1$ but $J_{23}>3$ in the example of the Appendix).

In \cite{altshuler-2009}, Altshuler et al.~claimed that the adiabatic quantum algorithm  with problem Hamiltonian $\ham_{AY}$ failed
with high probability for randomly generated instances of EC3.
They claimed that the correctness of their argument did not rely on the specific form of 
the problem Hamiltonian for Exact Cover, but only depended on the properties of the problem instance $B_i$ and $I_{ij}$.
However, our problem Hamiltonian  $\ham_{C}$ challenges the generality of their claim.
Their argument requires computing the energy difference $E_{12}(s)$ 
 which depends on the energy function of the problem Hamiltonian.
While the energy function for $\ham_{AY}$ only depends on $B_i$ and $I_{ij}$, the energy function for $\ham_C$
also depends on $J_{ij}$ (and/or $k$) whose values have a range to choose
\footnote{For example, using $\ham_{AY}$, the 2nd order correction
$E_x^{(2)} = -\sum_{i=1}^n 1/B_i$ (which results $E_{12}^{(2)}=0$).
But using  $\ham_C$,
we have
$
E_x^{(2)} = - (\sum_{\{i: x_i=0\}} 1/(B_i-\sum_{\{j\in nbr(i): x_j=1\}}J_{ij}) - \sum_{\{i: x_i=1\}}1/B_i )
$
which also depends on the flexible parameter $J_{ij}$ (assume $k=1$) and the neighbornood of $i$.
Here $E_{12}^2$ (and other higher term corrections)  depend on the connectivity of the graph, and the non-random choice of $J_{ij}$.
The argument in Altshuler et al. that ``$E_{12}^4$ is given by a sum of $\theta(N)$ random terms with zero mean'' no
longer applies here as $J_{ij}$ are not random.
}.

We would like to emphasize here we point out that the argument in \cite{altshuler-2009} does not carry over to this new adiabatic algorithm for the same problem (EC3). That means that we can not use their argument 
to claim that our new algorithm requires exponential time.
Whether this algorithm requires polynomial or exponential time will require rigorous analytical analysis of the algorithm. 
In \cite{young-2009}, Young~et~al. used QMC to show that $\gmin$ of adiabatic algorithm based on the same problem Hamiltonian $\ham_{AY}$ is exponentially small. 
It will be interesting to see  the $\gmin$ result (for the same set of instances) using this new problem Hamiltonian $\ham_{C}$.


\subsection{\SAT}

Similarly, for \SAT, there is a well-known reduction to MIS (which is one of the first NP-complete reductions, to show the NP-hardness of MIS)~\cite{garey-johnson}. For completeness, here we recall the reduction:

\paragraph{\SAT $\le_{P}$ MIS.} Given a \SAT instance $\Psi(x_1, \ldots, x_n) = C_1 \wedge \ldots \wedge C_m$ with $n$ variables and $m$ clauses, we construct a (unweighted) graph $\GSAT$ as follows:
\begin{itemize}
\item For each clause $C_i = y_{i_1} \vee y_{i_2} \vee y_{i_3}$, we construct a triangle with three vertices labeled  
accordingly, i.e., with $y_{i_1}, y_{i_2}, y_{i_3}$, where $y_j \in \{x_j, \bar{x_j}\}$. Therefore,  $\GSAT$ consists of $3m$ vertices.\item There is an edge between two vertices in different triangles if there labels are in conflict. That is, for $i \neq j$,   $i_sj_t \in \edge(\GSAT)$ if and only if $y_{i_s} = \bar{y_{j_t}}$. 
\end{itemize}
One can then show that $\Psi$ is satisfiable if and only if $\GSAT$ has a MIS of size $m$.
See e.g. \cite{DPV} for an example.




\section{Discussion}
\label{sec:discussion}

In this paper, 
we have shown that by 
changing the parameters in the problem Hamiltonian (without changing the problem to be solved) of the adiabatic algorithm for 
MIS on CK graphs, we prevent the FQPT, that causes the exponential small $\gmin$, from occurring and significantly increase $\gmin$.
We do so by scaling the vertex-weight of the graph, namely, multiplying the weights of vertices by a scaling factor.
In order to determine the best scaling factor, we raise the basic question about what the appropriate formulation of adiabatic running time should be. 

We also describe adiabatic quantum algorithms for \EC and \SAT  in which the problem Hamiltonians 
are based on the reduction to MIS. 
Notice that
the reduction requires only the solution  to be preserved,
i.e. there is 
 a polynomial time algorithm that maps
the solution to the 
reduced problem  to  the solution to the original problem and vice versa (see e.g. \cite{DPV}).
In other words, the reduction might only preserves the solution (i.e. the ground state) and alter the energy levels of the problem Hamiltonian. 
As we demonstrate in our small examples, the minimum spectral gap can be increased drastically when the excited energy levels are changed.
By definition, NP-complete problems can be polynomial reducible to each other.
Different reduction gives rise to different problem Hamiltonians, and thus different adiabatic algorithms, for the same problem.
At the risk of stating the obvious,
it is not sufficient to conclude that a problem is hard for adiabatic quantum computation/optimization 
by showing that there exists an adiabatic quantum algorithm 
 for the problem (e.g. for a particular problem Hamiltonian) that requires exponential time. 
There are three variable components, namely, initial Hamiltonian, problem Hamiltonian and evolution path, in order to specify an 
adiabatic algorithm. 
To prove that adiabatic quantum computation (optimization) fail to solve a particular problem in polynomial time, one requires to prove that
no polynomial-time adiabatic (optimization) algorithms for the problem is possible, which is in general hard.

In \cite{DMV01,DV}, van Dam et al.~argued that adiabatic quantum optimization might be thought of as a kind of ``quantum local search'', 
 and  in \cite{DV}, they constructed a special family of 3SAT instances for which the (clause-violation cost function based) adiabatic algorithm 
required exponential time\footnote{Farhi~et~al.~\cite{diff-path1} showed that the exponential small gap could be overcome by different initial Hamiltonians.}.
Our CK graphs were designed to trap local search algorithms
in the sense that there are many local minima to mislead the local search process.
From \desev{} on a 15-vertex CK graph,
 we see that indeed this is the case for $\ham_1$ and the adiabatic algorithm would require exponential time due to the
exponential small $\gmin$ caused by  the FQPT or the level anti-crossing between the global minimum and the local minima.
However, for $\ham_k$ (say $k=10$), the FQPT no longer occurs and $\gmin$ increases significantly, which might suggest the possibility of exponential speed-up
over $\ham_1$.
It remains challenging on how to analytically bound $\gmin$ and/or $\art$ of the adiabatic algorithm for $\ham_{k}$ on general
CK graphs. 
 


\section{Acknowledgments}
I would like to thank my very enthusiastic students in my adiabatic quantum computing class: 
Ryan Blace,
Russell Brasser, 
Mark Everline,
    Eric Franklin,
     Nabil Al Ramli,
  and   Aiman Shabsigh, who also helped to name \desev.
I would like to thank Siyuan Han and Peter Young for their comments. 
Thanks also go to David Sankoff and David Kirkpatrick for the encouragment.


\bibliographystyle{abbrv}

\begin{thebibliography}{10}

\setlength{\itemsep}{-0.5pt}

\bibitem{ADKLLR04}
D.~Aharonov, W.~van Dam, J.~Kempe, Z.~Landau, S.~Lloyd, and O.~Regev.
\newblock Adiabatic quantum computation is equivalent to standard quantum
  computation.
\newblock {\em SIAM Journal of Computing}, Vol. 37, Issue 1, p. 166--194 (2007), 
conference version in {\em Proc. 45th FOCS}, p. 42--51 (2004).




\bibitem{altshuler-2009}
B.~Altshuler, H.~Krovi and J.~Roland.
\newblock Adiabatic quantum optimization fails for random instances of
NP-complete problems.
\newblock arXiv:quant-ph/0908.2782, 2009.
\newblock Anderson localization casts clouds over adiabatic quantum optimization.
arXiv:quant-ph/0912.0746, 2009.

\bibitem{adt-AR}
A. Ambainis and O. Regev.
\newblock An elementary proof of the quantum adiabatic theorem.
\newblock arXiv:quant-ph/0411152.


\bibitem{AC09}  
M.H.S.~Amin, V.~Choi. 
\newblock First order phase transition in adiabatic quantum computation. 
\newblock arXiv:quant-ph/0904.1387, 2009. 
{\em Phys. Rev. A.}, {\bf 80}(6), 2009.

\bibitem{childs-clique}
A.M. Childs, E. Farhi, J. Goldstone and S. Gutmann.
\newblock Finding cliques by quantum adiabatic evolution.
\newblock {\em Quantum Information and Computation}, {\bf 2}, 181,
2002.


\bibitem{minor-embedding} 
V.~Choi. 
\newblock Minor-embedding in adiabatic quantum computation: I. The parameter setting problem.
\newblock {\em Quantum Inf. Processing.},  {\bf 7},  193--209, 2008.
Available at arXiv:quant-ph/0804.4884.


\bibitem{CK08}
V.~Choi, D.~Kirkpatrick. 
\newblock On the Construction of Hard Instances for the Maximum-Weight Independent Set Problem.
\newblock. Manuscipt. 2008.


\bibitem{Precision-scale}
V. Choi.
\newblock Scaling, Precision and Adiabatic Running Time.
\newblock In preparation.

\bibitem{primer}
V. Choi.
\newblock An Adiabatic Quantum Computation Primer.
\newblock In preparation.


\bibitem{DPV}
S. Dasgupta, C. Papadimitriou, U. Vazirani.
\newblock Algorihtms.
\newblock McGraw Hill, 2008.



\bibitem{FGGS00}
E.~Farhi, J.~Goldstone, S.~Gutmann, and M.~Sipser.
\newblock Quantum computation by adiabatic evolution.
\newblock arXiv:quant-ph/0001106, 2000.


\bibitem{FGGLLP01}
E.~Farhi, J.~Goldstone, S.~Gutmann, J.~Lapan, A.~Lundgren, and D.~Preda.
\newblock A quantum adiabatic evolution algorithm applied to random instances
  of an NP-complete problem.
\newblock {\em Science}, 292(5516):472--476, 2001.

\bibitem{diff-path1}
E. Farhi, J. Goldstone and S. Gutmann.
\newblock Quantum adiabatic evolution algorithms with different paths.
\newblock arXiv.org:quant-ph/0208135, 2002.

\bibitem{farhi-2009}
E. Farhi, J. Goldstone, D. Gosset, S. Gutmann, H. B. Meyer and P. Shor.
\newblock Quantum adiabatic algorithms, small gaps, and different paths.
\newblock arXiv.org:quant-ph/0909.4766, 2009.

\bibitem{garey-johnson}
M. R. Garey and D. S. Johnson. 
\newblock Computers and Intractability: A Guide to the Theory
of NP-Completeness. W. H. Freeman, 1979.

\bibitem{Hogg}
T.~Hogg.
\newblock Adiabatic quantum computing for random satisfiability problems.
\newblock {\em Phys. Rev. A} {\bf 67}, 022314, 2003.

\bibitem{jordan1}
S.P. Jordan, E. Farhi, P.W. Shor.
\newblock Error-correcting codes for adiabatic quantum computation.
\newblock {\em Phys. Rev. A.}, {\bf 74}, 052322, 2006.


\bibitem{Jorg1}
T.~Jorg, F.~Krzakala, G.~Semerjian, and F.~Zamponi.
\newblock First-order transitions for random optimization problems in a transverse field.
\newblock arXiv.org:quant-ph/0911.3438, 2009.


\bibitem{Jorg2}
T.~Jorg, F.~Krzakala, J. Kurchan, A.C. Maggs and J. Pujos.
\newblock Energy gaps in quantum first-order mean-field-like transitions:The 
problems that quantum annealing cannot solve.
\newblock arXiv.org:quant-ph/0912.4865, 2009.



\bibitem{Reichardt-04}
B.W. Reichardt.
\newblock The quantum adiabatic optimization algorithm and local minima.
\newblock {\em Proc. 35th STOC}, 502--510, 2004.

\bibitem{symmetries} 
G. Schaller and  R. Sch\"utzhold.
\newblock The role of symmetries in adiabatic quantum algorithms.
\newblock arXiv:quant-ph/0708.1882, 2007.

\bibitem{shor}
P.W.~Shor.
\newblock Algorithms for quantum computation: discrete logs and factoring.
\newblock {\em Proc. 35th FOCS}, (1994); {\em SIAM J. Comp.}, 26, 1484--1509, 1997.



\bibitem{DMV01}
W.~van Dam, M.~Mosca, and U.~Vazirani.
\newblock How powerful is adiabatic quantum computation?
\newblock {\em Proc. 42nd FOCS}, 279--287, 2001.

\bibitem{DV}
W.~van Dam and U.~Vazirani.
\newblock Limits on quantum adiabatic optimization.
\newblock {\em Unpublished}, 2001.

\bibitem{Young}  
A.P. Young, S. Knysh, and V.N. Smelyanskiy.
\newblock Size dependence of the minimum excitation gap in the quantum adiabatic algorithm.
\newblock {\em Phys. Rev. Lett.}, {\bf 101}, 170503, 2008.


\bibitem{young-2009}
 A.~P. Young and S. Knysh and V.~N. Smelyanskiy.
 First order phase transition in the Quantum Adiabatic Algorithm.
\newblock arXiv:quant-ph/0910.1378, 2009. 
\newblock {\em Phys. Rev. Lett.}, 2009.

\bibitem{young-note}
A.~P. Young.
\newblock Private Communication.

\bibitem{znidaric-2005-71}
M. Znidaric.
\newblock Scaling of running time of quantum adiabatic algorithm for
propositional   satisfiability.
\newblock {\em Phys. Rev. A}, {\bf 71}, 062305, 2005.



\begin{center}
  \textsc{Some Recent References on Adiabatic Theorem}
\end{center}




\bibitem{adt-1}
M.H.S. Amin.
\newblock On the inconsistency of the adiabatic theorem.
\newblock arXiv:quant-ph/0810.4335, 2008.  {\em Phys. Rev. Lett.} {\bf 102}, 220401, 2009. 

\bibitem{adt-2}
D. Comparat.
\newblock General conditions for quantum adiabatic evolution.
\newblock {\em Phys. Rev. A}, {\bf 80}, 012106, 2009.


\bibitem{adt-3}
V.I. Yukalov.
\newblock Adiabatic theorems for linear and nonlinear Hamiltonians.
\newblock {\em Phys. Rev. A}, {\bf 79}, 052117, 2009.


\bibitem{adt-4}
J. Du and L. Hu and Y. Wang and J. Wu and M. Zhao and D. Suter.
\newblock Is the quantum adiabatic theorem consistent?
\newblock arXiv:quant-ph/0810.0361, 2008.

\bibitem{adt-6}
J.~Goldstone.
\newblock Adiabatic Theorem.
\newblock Appendix F. S. Jordan's PhD Thesis. arXiv:quant-ph/0809.2307, 2008.


\bibitem{adt-5}
D.A. Lidar and A.T. Rezakhani and A. Hamma.
\newblock Adiabatic approximation with exponential accuracy for
many-body systems and quantum computation.
\newblock arXiv:quant-ph/0808.2697, 2008.











\bibitem{adt-9} 
 D.M. Tong, K. Singh, L.C. Kwek, and C.H. Oh.
\newblock Sufficiency Criterion for the Validity of the Adiabatic Approximation.
\newblock {\em Phys. Rev. Lett.} {\bf 98}, 150402, 2007. 

\bibitem{adt-10}
Z. Wei and M. Ying.
\newblock Quantum adiabatic computation and adiabatic conditions.
\newblock {\em Phys. Rev. A}, {\bf 76}, 024304, 2007.

\bibitem{adt-7}
Y. Zhao.
\newblock Reexamination of the quantum adiabatic theorem.
\newblock {\em Phys. Rev. A}, {\bf 76}, 032109, 2008.

\bibitem{adt-11}
R. MacKenzie, A. Morin-Duchesne, H. Paquette, and J. Pinel.
\newblock Validity of the adiabatic approximation in quantum mechanics.
\newblock {\em Phys. Rev. A}, {\bf 76}, 044102, 2007.

\bibitem{adt-8}
S. Jansen, R. Seiler and M.B. Ruskai.
\newblock Bounds for the adiabatic approximation with applications to
quantum computation.
\newblock {\em Journal of Mathematical Physics}, {\bf 48}, 102111, 2007.
Available at arXiv:quant-ph/0603175.



\end{thebibliography}

\newpage

{\bf\large Appendix A. }
  \paragraph{Example: \SEC $\le_{P}$ MIS.}

Let $\Psi(x_1, \ldots, x_7) = C_1 \wedge C_2 \wedge C_3 \wedge C_4 \wedge C_5$ be an instance of \SEC with $7$ variables and $5$ clauses:
\begin{itemize}
  \item  $C_1 = x_1 \vee x_2 \vee x_3$, $C_2 = x_1 \vee x_2 \vee x_4$, $C_3 = x_3 \vee x_4 \vee x_5$,
    $C_4 = x_1 \vee x_3 \vee x_6$, $C_5 = x_2 \vee x_6 \vee x_7$.
\end{itemize}

For each variable $x_i$, let $S_i$ be the set consisting of all clauses in which $x_i$ appears. That is, we have
\begin{itemize}
  \item $S_1=\{C_1,C_2,C_4\}$, $S_2=\{C_1,C_2,C_5\}$, $S_3=\{C_1,C_3,C_4\}$
    \item $S_4 = \{C_2, C_3\}$, $S_5=\{C_3\}$, $S_6=\{C_4,C_5\}$, $S_7=\{C_5\}$
\end{itemize}

\begin{figure}[hb]
  \begin{center}
\includegraphics[width=0.15\textwidth,angle=270]{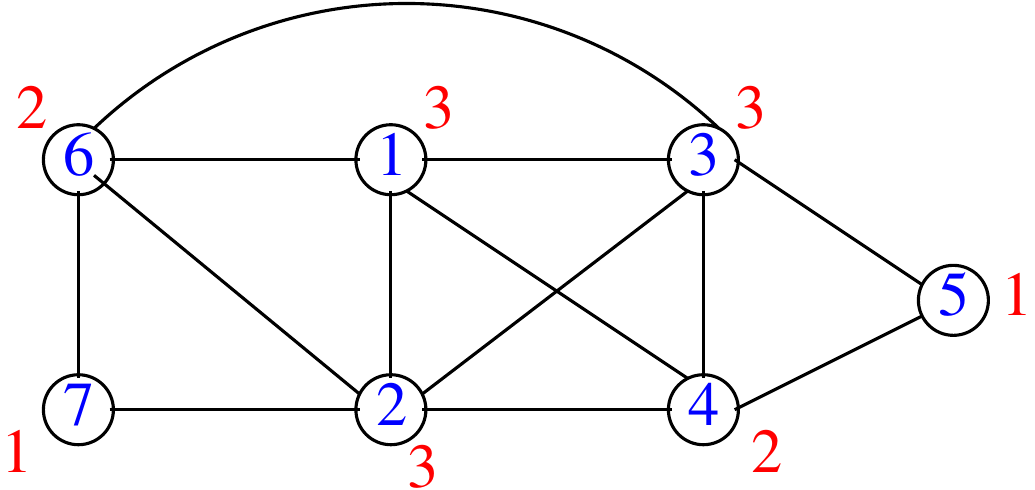}      
  \end{center}
\caption{The number next to the vertex is the weight of the vertex. $\wmis(\GEC)=\{1,5,7\}$, with weight 5.}
\label{fig:EC}
\end{figure}

Construct the graph $\GEC$ as follows:
\begin{itemize}
  \item $\ver(\GEC) = \{1, 2, \ldots, 7\}$, where vertex $i$ corresponds to the set $S_i$, and the weight of vertex $i$ is the number of elements in $S_i$ (=$B_i$);
\item $\edge(\GEC) = \{ij: S_i \mbox{ and } S_j$ share a common clause $\}$.
\end{itemize} 
See Figure~\ref{fig:EC}.
It is easy to see that $\GEC$ has a MIS of weight $5$ if only if $\Psi$ is satisfiable (in positive 1-in-3SAT sense).

\vspace*{-0.5cm}

  \begin{table}[h]
    \begin{center}
\begin{tabular}{|l|c|c|}
\hline
$w_B$ & $s^*$ & $\gmin$ \\ \hline
1.0&	0.2368&	5.23e-01\\ \hline
1.1&	0.2517&	4.12e-01\\ \hline 
1.2&	0.2708&	2.90e-01\\ \hline
1.3&	0.2964&	1.68e-01\\ \hline
1.4&	0.3323&	7.14e-02\\ \hline
1.5&	0.3805&	2.04e-02\\ \hline
1.6&	0.4422&	3.63e-03\\ \hline
1.7&	0.5217&	3.39e-04\\ \hline
1.8&	0.6276&	1.04e-05\\ \hline
1.9&	0.7758&	4.14e-08\\ \hline
\end{tabular}
    \end{center}
\caption{The minimum spectral gap $\gmin$ (and position $s^*$) changes as $w_B$ changes from $1$ to $1.9$, for the (unscaled) problem Hamiltonian $\ham_1$ in Eq.\eqref{eq:unscaled}.}
\label{table1}
\end{table}



\end{document}